\documentclass[fleqn,usenatbib]{mnras}

\usepackage{newtxtext,newtxmath}
\DeclareRobustCommand{\VAN}[3]{#2}
\let\VANthebibliography\thebibliography
\def\thebibliography{\DeclareRobustCommand{\VAN}[3]{##3}\VANthebibliography}

\usepackage{graphicx}	% Including figure files
\usepackage{amsmath}	% Advanced maths commands
\usepackage{pdflscape}
\usepackage{CJKutf8}
\usepackage{afterpage}

\newcommand{\pturb}{P_\mathrm{turb}}	% per cm-squared
\newcommand{\pbeamturb}{P_\mathrm{beam,turb}}
\newcommand{\pde}{P_\mathrm{DE}}
\newcommand{\avir}{\alpha_\mathrm{vir}}
\newcommand{\abeamvir}{\alpha_\mathrm{beam,vir}}
\newcommand{\aco}{\alpha_\mathrm{CO}}

\title[WISDOM Project -- XVII.\ Molecular Gas in Seven ETGs]{WISDOM Project -- XVII.\ Beam-by-beam Properties of the Molecular Gas in Early-type Galaxies}
\author[T.\ G.\ Williams et al.]{Thomas G.\ Williams,$^{1, 2}$\thanks{E-mail: thomas.williams@physics.ox.ac.uk (TGW)}
Martin Bureau,$^{1}$
Timothy A.\ Davis,$^{3}$
Michele Cappellari,$^{1}$
\newauthor
Woorak Choi,$^{4}$
Jacob S.\ Elford,$^{3}$
Satoru Iguchi,$^{5, 6}$
Jindra Gensior,$^{7}$
Fu-Heng Liang \begin{CJK*}{UTF8}{gbsn}(梁\,赋珩)\end{CJK*},$^{1}$
\newauthor
Anan Lu,$^{8}$
Ilaria Ruffa$^{3}$
and Hengyue Zhang \begin{CJK*}{UTF8}{gbsn}(张\,恒悦)\end{CJK*}$^{1}$
\\
$^{1}$Sub-department of Astrophysics, Department of Physics, University of Oxford, Keble Road, Oxford OX1 3RH, UK\\
$^{2}$Max Planck Institut f{\"u}r Astronomie, K{\"o}nigstuhl 17, 69117 Heidelberg, Germany\\
$^{3}$Cardiff Hub for Astrophysics Research \& Technology, School of Physics \& Astronomy, Cardiff University, Queens Buildings, The Parade, Cardiff, CF24 3AA, UK\\
$^{4}$Department of Astronomy, Yonsei University, 50 Yonsei-ro, Seodaemun-gu, Seoul 03722, Republic of Korea\\
$^{5}$Department of Astronomical Science, SOKENDAI (The Graduate University of Advanced Studies), Mitaka, Tokyo 181-8588, Japan\\
$^{6}$National Astronomical Observatory of Japan, National Institutes of Natural Sciences, Mitaka, Tokyo 181-8588, Japan\\
$^{7}$Institute for Computational Science, Winterthurerstrasse 190, 8057 Z\"{u}rich, Switzerland\\
$^{8}$Trottier Space Institute and Department of Physics, McGill University, 3600 University Street, Montreal, QC~H3A~2T8, Canada\\
}

\date{Accepted XXX. Received YYY; in original form ZZZ}

\pubyear{2022}

\begin{document}
\label{firstpage}
\pagerange{\pageref{firstpage}--\pageref{lastpage}}
\maketitle

% Abstract of the paper: 249/250 words
\begin{abstract}
We present a study of the molecular gas of seven early-type galaxies with high angular resolution data obtained as part of the mm-Wave Interferometric Survey of Dark Object Masses (WISDOM) project with the Atacama Large Millimeter/submillimeter Array. Using a fixed spatial scale approach, we study the mass surface density ($\Sigma$) and velocity dispersion ($\sigma$) of the molecular gas on spatial scales ranging from $60$ to $120$~pc. Given the spatial resolution of our data ($20$ -- $70$~pc), we characterise these properties across many thousands of individual sight lines ($\approx50,000$ at our highest physical resolution). The molecular gas along these sight lines has a large range ($\approx2$~dex) of mass surface densities and velocity dispersions $\approx40\%$ higher than those of star-forming spiral galaxies. It has virial parameters $\avir$ that depend weakly on the physical scale observed, likely due to beam smearing of the bulk galactic rotation, and is generally super-virial. Comparing the internal turbulent pressure ($\pturb$) to the pressure required for dynamic equilibrium ($\pde$), the ratio $\pturb$/$\pde$ is significantly less than unity in all galaxies, indicating that the gas is not in dynamic equilibrium and is strongly compressed, in apparent contradiction to the virial parameters. This may be due to our neglect of shear and tidal forces, and/or the combination of three-dimensional and vertical diagnostics. Both $\avir$ and $\pturb$ anti-correlate with the global star-formation rate of our galaxies. We therefore conclude that the molecular gas in early-type galaxies is likely unbound, and that large-scale dynamics likely plays a critical role in its regulation. This contrasts to the giant molecular clouds in the discs of late-type galaxies, that are much closer to dynamical equilibrium.
\end{abstract}


%%%%%%%%%%%%%%%%%%%%%%%%%%%%%%%%%%%%%%%%%%%%%%%%%%

%%%%%%%%%%%%%%%%% BODY OF PAPER %%%%%%%%%%%%%%%%%%

\section{Introduction}\label{sec:intro}

The majority of massive (\textgreater~$10$~$\mathrm{M_\odot}$) star formation occurs within dense \citep[mass surface densities \textgreater~$50$~$\mathrm{M_\odot~pc^{-2}}$; e.g.][]{1987Solomon, 2010Hughes}, compact \citep[tens of parsecs; e.g.][]{2010RomanDuval,2017MivilleDeschenes} giant molecular clouds \citep[GMCs;][]{2003LadaLada}. As the fundamental unit for star formation, and hence galaxy growth, an understanding of the processes driving the formation and collapse of these molecular clouds is vital to ultimately understand galaxy evolution. 

Given the compact nature of GMCs, studies have until recently been limited to the Milky Way (MW) and Local Group galaxies. Early studies in the MW led to a series of three relationships known widely as the \cite{1981Larson} relations\footnote{Or Larson's laws. We avoid this terminology here, as these relations have been derived empirically rather than from theoretical considerations.}, whereby GMCs tend to have roughly constant molecular gas mass surface densities ($\Sigma$), be in virial equilibrium and follow a size ($R$) -- linewidth ($\sigma$) relation (i.e.\ larger clouds have higher velocity dispersions). \cite{2009Heyer} showed that these three relationships are linked, and can be visualised within a single plane ($\sigma/R^{0.5}$ versus $\Sigma$, often referred to as the `Heyer plot'). If all three Larson relations are obeyed, GMCs will lie at a single position in this plane. Using a less biased sample than that of \cite{1981Larson}, \cite{2009Heyer} however showed that clouds typically follow a linear relation in this plane, such that clouds are roughly in virial equilibrium and obey the size -- linewidth relation but have a range of mass surface densities. This appears to be borne out by other studies of Local Group galaxies \citep[e.g.][]{2003Rosolowsky,2007Rosolowsky}.

With the advent of the Atacama Large Millimeter/submillimeter Array (ALMA), a new window on the cold molecular gas in the nearby Universe has been opened up. ALMA can obtain GMC-scale observations of statistically-complete samples of galaxies, finally allowing robust studies of molecular clouds across the galaxy population. With such observations, it is possible to study how cloud properties vary as a function of e.g.\ galaxy morphology and/or environment. For star-forming, late-type galaxies (LTGs), the Physics at High Angular resolution in Nearby GalaxieS\footnote{\url{phangs.org}} \citep[PHANGS;][]{2021bLeroy} survey provides the largest homogeneous study of cloud properties. It suggests that clouds are similar to those of the MW across the galaxies probed, with the exception of the centres of barred galaxies, that have elevated velocity dispersions \citep{2018Sun, 2020Sun,2021Rosolowsky}. Taken together, this paints a picture of clouds being long-lived stable structures, relatively homogeneous across the LTG population, and relatively insensitive to the large-scale dynamical structures they reside in (with the exception of the much more dynamically-active galaxy centres).

However, ALMA also allows to look beyond main-sequence star-forming galaxies, where the picture is less clear and much more poorly understood. In ultra-luminous infrared galaxies (ULIRGs), that often have star-formation rates (SFRs) significantly higher than those of main-sequence galaxies, the velocity dispersions of clouds are significantly smaller than those of equivalently-sized clouds in the MW (Saito et al., in preparation). For early-type galaxies \citep[ETGs; often referred to as `red and dead' galaxies due to a lack of star formation; see][]{2014Davis}, molecular gas is detected in about a quarter of objects \citep{2011Young}. The availability of gas but lack of star formation is puzzling. `Morphological' or `dynamical' quenching is postulated as the most likely culprit \citep{2009Martig, Jeffresoninprep}, whereby gas becomes stable against collapse through the growth of a stellar bulge. Other works have posited that the lack of star formation is due to the unavailability of cold gas \citep{2017Kokusho}. Whatever the reason, the cloud properties are significantly different from those of star-forming galaxies. For example, the velocity dispersions are typically (but not always) significantly higher \citep[e.g.][]{2015Utomo, 2017Davis}. Critically, the clouds of ETGs may not be in virial equilibrium, and thus none of the Larson relations may hold \citep{2021Liu}. 

In this work, we make use of state-of-the-art ALMA observations of seven ETGs from the mm-Wave Interferometric Survey of Dark Object Masses\footnote{\url{https://wisdom-project.org/}} (WISDOM) project, to provide a more holistic overview of the molecular clouds and their properties in these poorly-studied galaxies. We use those data to study, on a beam-by-beam basis, the molecular gas properties of our sample ETGs, and to explore the differences (if any) between the state of molecular gas in ETGs and that in star-forming main-sequence galaxies. Whilst WISDOM was initially conceived to measure supermassive black hole masses through molecular gas kinematic modelling \citep[e.g.][]{2013Davis, 2019North, 2020Davis, 2021Smith, 2023Ruffa}, the data obtained necessarily have high spatial resolutions (to probe the spheres of influence of the black holes), and so are ideal for molecular cloud-scale studies. The simplicity of the beam-by-beam technique allows to study samples larger than those previously considered in the literature, and it offers a simple framework for comparison to simulations. The ETG sample considered here is significantly different from the LTG samples that have been studied on a beam-by-beam or cloud-by-cloud basis previously \citep[e.g.][]{2010Hughes, 2018Sun, 2021Rosolowsky}.

The layout of this paper is as follows. In Section~\ref{sec:data}, we present an overview of the data used, as well as the reduction procedures adopted to produce spectral-line cubes. In Section~\ref{sec:pix_by_pix_measurements}, we detail our methodology to extract measurements of molecular gas properties, and compare these measurements to the Larson relations. We also calculate the internal turbulent pressures, and compare these to expectations from dynamical equilibrium. We discuss the implications of our results in Section~\ref{sec:discussion}, and summarise our main results in Section~\ref{sec:conclusions}. Throughout this work, stellar masses and star formation rates have been derived assuming a \cite{2001Kroupa} initial mass function (IMF).

\section{Data Overview and Preparation}\label{sec:data}

Our sample is a subset of galaxies mapped in low-{\it J} CO transitions with ALMA by the WISDOM survey. We only select ETGs from WISDOM, as the full sample also includes LTGs and dwarfs \citep[see the Hubble types in][]{2022Davis}. These are classified as ETGs either due to their membership in ATLAS$^{\rm 3D}$ \citep{2011Cappellari}, or for galaxies not present in ATLAS$^{\rm 3D}$ from visual inspection of {\it Hubble Space Telescope} ({\it HST}) images by \cite{2022Davis}. We then further select only galaxies with a multi-Gaussian expansion \citep[MGE;][]{1994Emsellem,2002Cappellari} model of the stellar mass distribution, allowing to calculate the dynamical pressure in Section~\ref{sec:pressure}. For this we search both earlier WISDOM works (as the stellar mass distribution of a galaxy is required to calculate its black hole mass) and the ATLAS$^{\rm 3D}$ survey \citep{2011Cappellari} mass models of \citet{2013Scott}. This leaves us with a total of seven galaxies\footnote{\citet{2023Ruffa} also present a MGE model of the galaxy NGC~4261, but given the poor velocity resolution of the compact configuration 12-m data ($5$~km~s$^{-1}$ versus the $2.5$~km~s$^{-1}$ we target), we exclude this galaxy.}, whose fundamental parameters are listed in Table~\ref{tab:data_overview}. For galaxies that have MGE models from both WISDOM and ATLAS$^{\rm 3D}$, we adopt the WISDOM models, as they are fitted to higher spatial resolution {\it HST} images. The galaxies with ATLAS$^{\rm 3D}$ MGE models are NGC~3607 and NGC~4435; the other galaxies all have MGE models from WISDOM works, that improve on these earlier models by using higher-resolution {\it HST} data rather than SDSS imaging. We have tested the difference between the ATLAS$^{\rm 3D}$ and WISDOM MGE models for NGC~4697, finding a mass ratio between the two of $1.0^{+0.3}_{-0.2}$ within the central $20\arcsec$ region, indicating a generally good agreement. WISDOM MGEs often have a number of more centrally peaked components (reflecting the higher resolution of the {\it HST} data).

\begin{table*}
\caption{Overview of the galaxies used in this study. \label{tab:data_overview}}
\begin{tabular}{cccccccc}
Galaxy & Distance$^\mathrm{a}$ & $T_\mathrm{Hubble}^\mathrm{b}$ & $\log\left(M_\ast/\mathrm{M}_\odot\right)^\mathrm{c}$ & $\log\left(\mathrm{SFR}/M_\odot~{\rm yr}^{-1}\right)^\mathrm{d}$ & $R_\mathrm{e}^\mathrm{e}$ & Native resolution & MGE Ref. \\
 & (Mpc) & & & & (arcsec) & (pc) & \\
\hline
NGC0383 & 66.6 & $-2.9\pm0.6$ & $11.8\pm0.1$ & $\phantom{-}0.001\pm0.2$ & 11\phantom{.11} & 62.4 & \cite{2019North}\\
NGC0524 & 23.3 & $-1.2\pm0.6$ & $11.4\pm0.1$ & $-0.56\phantom{1}\pm0.2$ & 23.66 & 51.7 & \cite{2019Smith}\\
NGC1574 & 19.9 & $-2.9\pm0.5$ & $10.8\pm0.1$ & $-0.1\phantom{11}\pm0.2$ & 21.01 & 20.8 & \cite{2023Ruffa}\\
NGC3607 & 22.2 & $-3.2\pm1.4$ & $11.3\pm0.1$ & $-0.54\phantom{1}\pm0.2$ & 21.9\phantom{1} & 70.5 & \cite{2013Scott}\\
NGC4429 & 16.5 & $-0.8\pm1.5$ & $11.2\pm0.1$ & $-0.84\phantom{1}\pm0.2$ & 48.84 & 16.1 & \cite{2018Davis}\\
NGC4435 & 16.7 & $-2.1\pm0.5$ & $10.7\pm0.1$ & $-0.84\phantom{1}\pm0.2$ & 28.49 & 55.4 & \cite{2013Scott}\\
NGC4697 & 11.4 & $-4.5\pm0.8$ & $11.1\pm0.1$ & $-1.08\phantom{1}\pm0.5$ & 39.51 & 36.2 & \cite{2017Davis}\\
\hline
\end{tabular}
\\Notes: (a) \citet{2017Steer}. (b) Numerical Hubble type \citep[HYPERLEDA;][]{2014Makarov}. (c) Total stellar mass: NGC0383 \citep[MASSIVE;][]{2014Ma}, NGC0524, NGC1574 \citep[z0MGS;][]{2019Leroy}, NGC3607, NGC4429, NGC4435 and NGC4697 \citep[ATLAS$^{\rm 3D}$;][]{2011Cappellari}. (d) Star-formation rate compiled by \citet{2022Davis} (from \citealt{2014Davis} and \citealt{2016Davis}) and \cite{2019Leroy}. (e) Effective radius \citep[2MASS;][]{2000Jarrett}. Distances and effective radii are published in the original works without uncertainties, so none is tabulated here. Stellar masses and star-formation rates are dominated by the instrument calibration uncertainties, so all measurement uncertainties are typically the same. 
\end{table*}

Given that the PHANGS survey presents a clear point of comparison for our study, and given the reasonably similar physical resolutions and choice of CO transition, we reduced the WISDOM data in a manner completely analogous to that of the PHANGS data. We use the PHANGS-ALMA processing pipeline\footnote{With some code modifications to allow for the larger data volumes of the higher angular resolution WISDOM data compared to those of the PHANGS-ALMA data. The code is available in the public version of the pipeline at \url{https://github.com/akleroy/phangs_imaging_scripts/}.} \citep{2021aLeroy} and all the data (i.e.\ all the tracks) available for our targets.
% \footnote{For NGC~0383, these are ALMA programmes 2012.1.01092.S, 2015.1.00419.S, 2016.1.00437.S and 2016.2.00053.S. For NGC~0524, 2015.1.00466.S, 2016.2.00053.S and 2017.1.00391.S. For NGC~1574, 2015.1.00419.S and 2016.2.00053.S. For NGC~3607, 2015.1.00598.S and 2016.2.00053.S. For NGC~4429, 2013.1.00493.S. For NGC~4435, 2015.1.00598.S and 2016.2.00053.S. For NGC~4697, 2015.1.00598.S.}
We note that this generally includes both the main ALMA 12-m array as well as the lower resolution 7-m Atacama Compact Array (ACA), but not total power data. This is unlikely to be a concern for the overall absolute flux levels as the maximum extents of the CO discs are much smaller than the maximum recoverable scales of the ACA observations. The exceptions are NGC~4429 and NGC~4697, which only have 12-m array data (but are reasonably compact in CO).

The PHANGS-ALMA pipeline is discussed at length in \cite{2021aLeroy}, and we refer readers to that work for details. Briefly, the data from the various arrays are first continuum-subtracted in the $uv$ plane, before being concatenated into a single measurement set (MS). A pixel scale is chosen to oversample the beam by a factor of around seven, and the image size is chosen to cover the field of view, whilst also being an optimal number of pixels for the fast-Fourier transforms the imaging uses. Initially, a shallow multi-scale {\tt CLEAN} \citep{2008Cornwell} is performed, which is optimised for extended sources. However, deep {\tt CLEAN}ing with this algorithm can produce spurious artefacts such as negative bowls, so this is only performed to a level of four times the root-mean-square (RMS) noise measured from the initial dirty cube. Following this, the more standard \cite{1974Hogbom} single-scale {\tt CLEAN} is used, {\tt CLEAN}ning down to the RMS noise (or until the changes in the residual images are negligible). In both {\tt CLEAN} stages, a {\tt CLEAN} mask is used -- it is less restrictive in the multi-scale step, and more restrictive in the single-scale step. The synthesised beam is then circularised, as a typically elliptical beam complicates the simple beam-by-beam analysis. To ensure the synthesised beam is critically sampled but minimise the data volume, the data are then rebinned so that the beam is oversampled by a factor of only about three. Any extra padding of the data is removed to optimise the image size. At this stage, we also produce a number of cubes at fixed spatial resolutions of $60$, $90$ and $120$~pc (if the spatial resolution of the data allows for this), by spatially convolving the native resolution cubes with an appropriate Gaussian kernel. We note that one should formally taper the $uv$-data rather than spatially convolve the final cube, but tests show that these two approaches lead to small differences ($<10\%$; \citealt{2022Davis}) for moderate convolutions, so we are confident that our chosen method does not bias our results. It also allows to directly compare to PHANGS results. The cubes are then binned in velocity to a channel width of $2.5$~km~s$^{-1}$, which is close to the native channel width of the data ($\approx1$~km~s$^{-1}$) and allows to resolve the typical line widths of molecular clouds \citep[a few~km~s$^{-1}$; e.g.][]{2020Sun}.

\begin{figure*}
	\includegraphics[width=\textwidth]{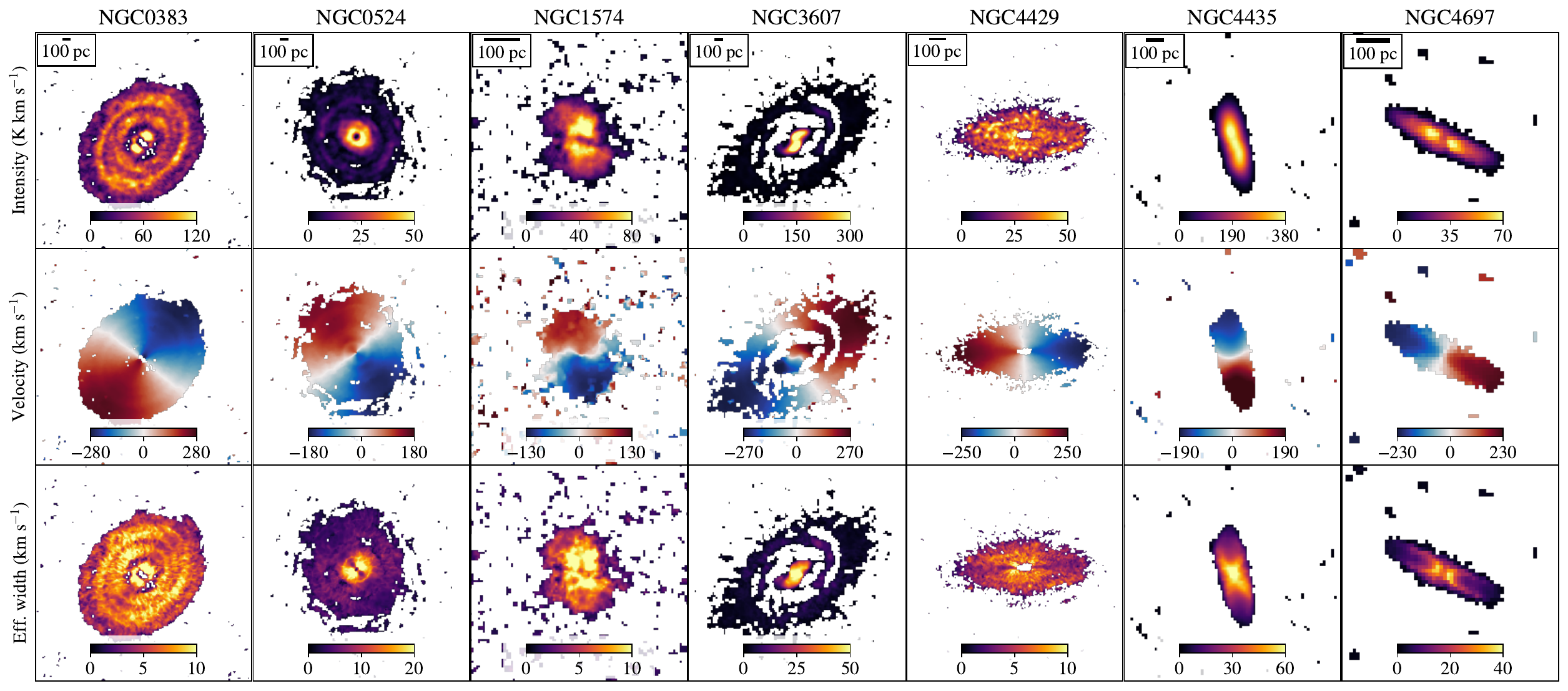}
    \caption{Overview of the data used in this study. Galaxies are ordered alphabetically from left to right. From top to bottom for each galaxy, the panels show the integrated intensity map (in K~km~s$^{-1}$), luminosity-weighted mean velocity map (in km~s$^{-1}$) and luminosity-weighted effective width map (in km~s$^{-1}$) at the native angular resolution. For each galaxy, a $100$~pc scale bar is shown in the top-left corner of the top panel.}
    \label{fig:data}
\end{figure*}

To ensure we consider only bona-fide emission in our analysis, we use `strict' masks for the cubes, that include high-confidence emission at the expense of slightly lower completeness. This mask creation essentially follows the recipe of \cite{2006RosolowskyLeroy}, whereby regions above a signal-to-noise ratio (S/N) threshold of $4$ are expanded down in S/N (spatially and spectrally) to regions above a S/N of $2$. For a quantitative comparison of `strict' versus `broad' masking and their performance, we refer the reader to \cite{2021aLeroy}. Using these masks, we create standard products for the ALMA data: integrated-intensity (moment~0), luminosity-weighted mean velocity (moment~1) and luminosity-weighted effective width \citep[a non-parametric estimate of the line width; see equation A7 of][]{2001Heyer} maps. An overview of these derived products is shown in Figure~\ref{fig:data}. We note that these maps look somewhat different from the analogous maps shown in earlier WISDOM works. The reasons for this are two-fold. Firstly, the data used here may include more observational tracks (especially the inclusion of ACA data). Secondly, our data processing and strict masking can lead to somewhat different results. We compared our cubes to those of these earlier works and found the flux level differences to be at most $\approx1\%$ \citep[see][]{2019North, 2019Smith, 2017Davis, 2018Davis}. This provides confidence that our choice of processing does not affect our results.

\section{Beam-by-beam Measurements}\label{sec:pix_by_pix_measurements}

Rather than attempting to perform a complex cloud extraction on our data, we take a non-parametric approach, to calculate `beam-by-beam' measurements of the molecular gas properties. Such an approach has been advocated by \cite{2016Leroy}, as it is easy to apply to large datasets and involves minimal assumptions. In a traditional cloud-extraction approach, different cloud extraction algorithms lead to identifying different cloud populations, even when attempting to match parameters between them \citep{2019Williams}. Our adopted beam-by-beam analysis does not measure cloud properties, as no cloud is ever identified, but it is broadly related to the more traditional cloud extraction methods. It is also significantly simpler to apply to large samples of galaxies, and has utility in characterising all of the gas in a galaxy (whilst cloud extraction approaches usually do not include all of the emission in the segmentation process). The two approaches nevertheless do appear to produce similar results \citep{2018Sun, 2020Sun, 2021Rosolowsky}, and both are valid if conceptually different.

The two main quantities we focus on in this work are the molecular gas mass surface density and the velocity dispersion. For the former, we use the integrated-intensity map (see Section~\ref{sec:data}) and apply a standard MW CO-to-molecule (i.e.\ including both molecular hydrogen and contributions from heavier elements) conversion factor
\begin{equation}\label{eq:co_conv}
    \alpha_\mathrm{CO(1-0)}=\frac{4.35}{R_{21}}\,\mathrm{M_\odot~pc^{-2}~\left(K~km~s^{-1}\right)}^{-1}\,,
\end{equation}
where $R_{21}$ is the CO ({\it J}=2-1)/({\it J}=1-0) ratio of integrated fluxes expressed in brightness temperature units, and is assumed to be $0.7$ here \citep[a suitable ratio across a large range of galaxies; e.g.][]{2021DenBrok, 2022Leroy}. We note that the CO conversion factor and line ratio are uncertain in ETGs, but given the approximately solar metallicity of the gas we do not expect large variations of these factors (likely within factors of a few). We discuss this uncertainty further in Section~\ref{sec:alpha_co_variation}. Given the high spatial resolution of our observations, we do not apply any beam-filling correction factors.

As a proxy for the true line width, we use the \cite{2001Heyer} effective width
\begin{equation}\label{eq:eff_width}
    \sigma_\mathrm{EW,measured}=\frac{I_\mathrm{CO}}{\sqrt{2\pi}\,T_\mathrm{peak}}\,,
\end{equation}
where $I_\mathrm{CO}$ is the integrated intensity and $T_\mathrm{peak}$ the peak intensity along each sight line. We use the subscript EW throughout to denote this quantity, but note that it is different from the optical definition of a line `equivalent width' (and from the usual definition of velocity dispersion). This definition is calibrated to return the centred second moment for a perfectly Gaussian line-of-sight velocity distribution (LOSVD), but it does not require the LOSVD to be necessarily Gaussian. It also has the benefit of being less sensitive to noise, and is free from assumptions about the shape of the LOSVD (and is thus less affected by potential multiple peaks along the line of sight). Because of the finite channel width of data, this measurement becomes increasingly inaccurate for line widths similar to or smaller than the velocity resolution of the data. We correct for this by deconvolving the spectral response following \cite{2006RosolowskyLeroy}:
\begin{equation}\label{eq:vel_deconv}
    \sigma_\mathrm{EW}=\sqrt{\sigma_\mathrm{EW, measured}^2-\sigma_\mathrm{response}^2}\,,
\end{equation}
where
\begin{equation}\label{eq:sigma_resp}
    \sigma_\mathrm{response}\approx\frac{\Delta v}{\sqrt{2\pi}}\times(1+1.18k+10.4k^2)
\end{equation}
\citep{2016Leroy} and
\begin{equation}\label{eq:resp_k}
    k\approx0+0.47r-0.23r^2-0.16r^3+0.43r^4\,,
\end{equation}
where $\Delta v$ is the channel width (here $2.5$~km~s$^{-1}$) and $r$ is the correlation measured from successive line-free channels. The correction to $\sigma_\mathrm{EW, measured}$ is generally quite small, but again it becomes significant for $\sigma_\mathrm{EW, measured}$ similar to or smaller than our velocity resolution.

We use the uncertainty maps generated by the PHANGS-ALMA pipeline, that are based on a Gaussian propagation of uncertainties. Formally, the uncertainties on $\Sigma$ and $\sigma_\mathrm{EW}$ are correlated with each other (as $\sigma_\mathrm{EW}$ is defined by $I_\mathrm{CO}$), but using Monte-Carlo methods to propagate this covariance more rigorously leads to negligible changes in the measured relationships between these quantities, at the expense of a significantly higher computation time \citep{2016Leroy, 2018Sun}. We note that we use exactly the same definitions of the measured quantities as \cite{2018Sun} and \cite{2020Sun} for the PHANGS data, so our results are directly and immediately comparable. Across our sample, we find a median velocity dispersion of $\approx5$~km~s$^{-1}$ at $90$~pc resolution (the minimum spatial resolution that includes all of our sample galaxies), similar to the velocity dispersions measured in earlier cloud studies of ETGs at higher spatial resolutions \citep[e.g.][]{2015Utomo, 2021Liu}. This provides confidence that (at least at our higher resolutions) beam smearing of the ordered rotation of the galaxies is not significantly affecting our $\sigma_\mathrm{EW}$ measurements.

\subsection{The \texorpdfstring{$\sigma_\mathrm{EW}$ -- $\Sigma$}{sigma-Sigma} relationship}\label{sec:sigma_sigma}

An isolated self-gravitating distribution of particles in steady state satisfies the virial theorem, $K=-W/2$, where $K$ is the total kinetic energy and $W$ the total gravitational potential energy \citep[e.g.][section~4.3]{1987BinneyTremaine}. For a spherical distribution with total mass $M$, radius $R$ and uniform mass volume density, $W=-3GM^2/5R$, where $G$ is the gravitational constant. In general $K=M\langle v^2\rangle/2$, where $\langle v^2\rangle$ is the mean squared velocity of the particles. In spherical symmetry, $\langle v^2\rangle=3\,\sigma_\mathrm{RMS}$, where $\sigma_\mathrm{RMS}$ is the second moment of the LOSVD, namely the line-of-sight velocity dispersion. Inserting these expressions in the virial formula, one obtains
\begin{equation}\label{eq:heyer_sigma_sigma}
    \sigma_\mathrm{RMS}=\left(\frac{\pi G}{5}\right)^{1/2}\,\Sigma^{1/2}\,R^{1/2}\,.
\end{equation}
We follow \cite{2018Sun} by taking the logarithm of this equation, allowing for a free exponent $\beta$ for $\Sigma$ and isolating the radius and the normalisation coefficient into the free parameter $A$, that therefore represents the normalisation at $100$~$\mathrm{M_\odot~pc^{-2}}$:
\begin{equation}\label{eq:sun_sigma_sigma}
    \log_{10}\left(\frac{\sigma_\mathrm{RMS}}{\mathrm{km~s^{-1}}}\right)=\beta\log_{10}\left(\frac{\Sigma}{\mathrm{100~M_\odot~pc^{-2}}}\right)+A.
\end{equation}

This equation formally applies only to clouds in virial equilibria, of line-of-sight velocity dispersion $\sigma_\mathrm{RMS}$, mass surface density $\Sigma$ and size $R$. To interpret the results of our beam-by-beam analysis, we therefore make two key assumptions: i) the synthesised beam size is similar to the size of the clouds in the galaxy observed and ii) the beam is filled with bright CO emission (i.e.\ clouds). The first assumption is specific to this method, but our native resolutions of tens of parsecs, and fixed spatial scales, are all of the same order as the sizes of MW \citep[e.g.][]{2010RomanDuval, 2017MivilleDeschenes} and ETG \citep[e.g.][]{2021Liu} clouds, so it should be reasonable. The beam dilution effect underlying the second assumption is common to all methods where clouds are only moderately spatially resolved. Given the high spatial resolutions of our data, we expect that this second assumption should also hold, although there may be beams with little or no gas detected along the sight line. In this sense, the relationships probed throughout this work are overall averages of the gas properties of the galaxies, but any individual measurement may not be robust. The fact that the traditional cloud analysis \citep[e.g.][]{2021Rosolowsky} and the beam-by-beam approach \citep[e.g.][]{2018Sun} tend to reach the same conclusions bears this out.

Assuming $\sigma_\mathrm{RMS}\approx\sigma_\mathrm{EW}$ in our beam-by-beam analysis, for identical clouds in virial equilibria the relation should have $\beta=1/2$ and $A$ a constant that depends on the synthesised beam size. For each galaxy at each spatial scale (native, $60$, $90$ and $120$~pc\footnote{In practice, if the native resolution of the data is within $10\%$ of the spatial scale being fit, we use those data as they are. This acknowledges the fact that there are uncertainties in the galaxy distances adopted.}), we fit for $\beta$ and $A$ using a Monte-Carlo Markov chain (MCMC) analysis utilising {\tt emcee} \citep{2013ForemanMackey}. We also include an intrinsic scatter term ($\Delta$), that is generally smaller than the statistical uncertainties on the individual measurements. This indicates that the scatter between the various sight lines is simply due to the uncertainties of our measured quantities. Throughout this work, we do not take distance uncertainties into account, as this would only lead to a systematic shift of the relation (i.e.\ $A$) for each galaxy but not affect the slope (i.e.\ $\beta$). 

The best-fitting parameters of each galaxy are listed Table \ref{tab:sigma_sigma_fits}. We show an example of the $\sigma_\mathrm{EW}$ -- $\Sigma$ relation for the galaxy NGC~4429 in Fig.~\ref{fig:sigma_sigma_ngc4429}. Analogous figures for the other sample galaxies are shown in Appendix~\ref{app:sigma_sigma}. There is a strong correlation and thus a clear relation between $\sigma_\mathrm{EW}$ and $\Sigma$, with a \cite{Kendall1938} robust rank correlation coefficient $\tau$ greater than $0.5$ at all physical scales. At a given molecular gas mass surface density, the line widths are elevated relative to the expectation from virial equilibrium, indicating a virial parameter $\alpha_\mathrm{vir}$ greater than $1$.

\begin{figure*}
	\includegraphics[width=\textwidth]{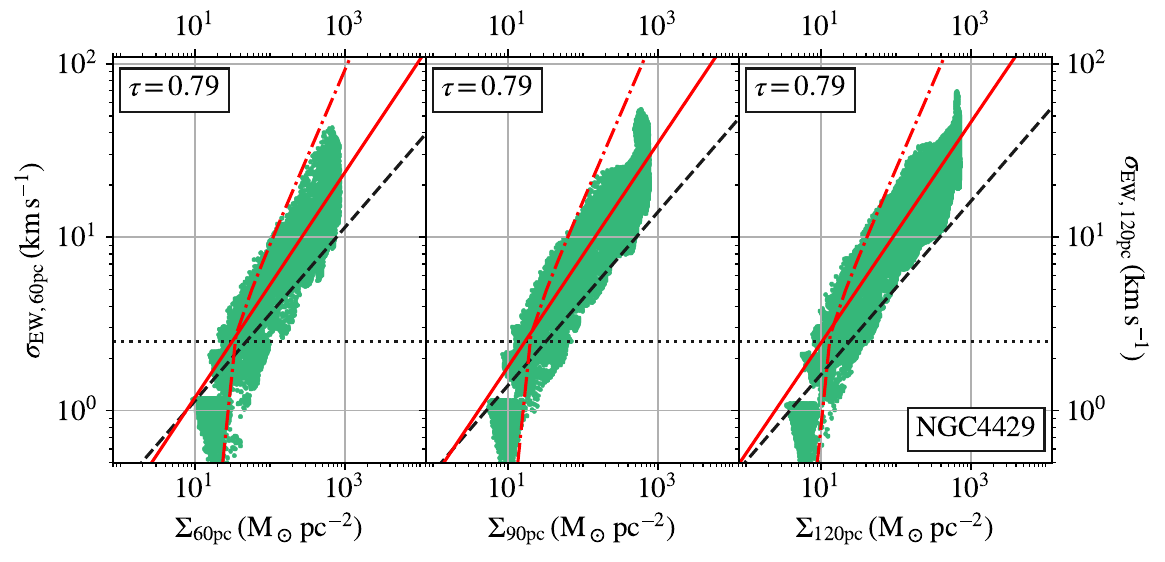}
    \caption{$\sigma_\mathrm{EW}$ -- $\Sigma$ relationship of the galaxy NGC~4429, at a number of spatial resolutions. In each panel, the dot-dashed red line shows our 3$\sigma$ completeness limit \citep[calculated using the method described in appendix~C of][]{2020Sun}. We only fit to data points at larger $\Sigma$. The solid red line shows the best-fitting relation following Eq.~\ref{eq:sun_sigma_sigma}; the uncertainties on the fit are of order the thickness of the line. The black dashed line shows the fiducial virial equilibrium relation. The black horizontal dotted line is the channel width of our data. We list the Kendall-$\tau$ correlation coefficient of the data points in the top-left corner. The colour of the data points matches that of Figure~\ref{fig:sigma_sigma}. Analogous plots for all the galaxies are shown in Appendix~\ref{app:sigma_sigma}.}
    \label{fig:sigma_sigma_ngc4429}
\end{figure*}

\afterpage{
\begin{landscape}
\begin{table}
\caption{Best-fitting parameters of the $\sigma$ - $\Sigma$ relation of each galaxy (see Eq.~\ref{eq:sun_sigma_sigma}). The subscript of each parameter indicates the spatial resolution at which the fit was performed. \label{tab:sigma_sigma_fits}}
\begin{tabular}{cccccccccc}
Galaxy & $\beta_{\rm 60pc}$ & $A_{\rm 60pc}$ & $\Delta_{\rm 60pc}$ & $\beta_{\rm 90pc}$ & $A_{\rm 90pc}$ & $\Delta_{\rm 90pc}$ & $\beta_{\rm 120pc}$ & $A_{\rm 120pc}$ & $\Delta_{\rm 120pc}$\\
\hline
NGC0383 & $0.705\pm0.014$ & $0.410\pm0.008$ & $0.002\pm0.002$ & $0.630\pm0.008$ & $0.583\pm0.004$ & $0.001\pm0.001$ & $0.601\pm0.006$ & $0.690\pm0.003$ & $0.022\pm0.005$\\
NGC0524 & $0.549\pm0.005$ & $0.985\pm0.003$ & $0.123\pm0.001$ & $0.550\pm0.004$ & $1.120\pm0.003$ & $0.138\pm0.001$ & $0.562\pm0.004$ & $1.220\pm0.003$ & $0.146\pm0.001$\\
NGC1574 & $0.817\pm0.009$ & $0.803\pm0.003$ & $0.058\pm0.004$ & $0.794\pm0.007$ & $0.966\pm0.003$ & $0.072\pm0.003$ & $0.782\pm0.006$ & $1.089\pm0.002$ & $0.080\pm0.002$\\
NGC3607 & -- & -- & -- & $0.782\pm0.005$ & $0.875\pm0.002$ & $0.104\pm0.002$ & $0.787\pm0.006$ & $0.990\pm0.002$ & $0.126\pm0.002$\\
NGC4429 & $0.649\pm0.003$ & $0.726\pm0.001$ & $0.131\pm0.001$ & $0.645\pm0.002$ & $0.897\pm0.001$ & $0.143\pm0.001$ & $0.637\pm0.002$ & $1.026\pm0.001$ & $0.147\pm0.001$\\
NGC4435 & $0.597\pm0.012$ & $0.885\pm0.011$ & $0.100\pm0.006$ & $0.577\pm0.011$ & $1.059\pm0.009$ & $0.139\pm0.005$ & $0.536\pm0.011$ & $1.179\pm0.008$ & $0.140\pm0.004$\\
NGC4697 & $0.786\pm0.022$ & $1.132\pm0.009$ & $0.161\pm0.007$ & $0.797\pm0.015$ & $1.301\pm0.007$ & $0.134\pm0.005$ & $0.773\pm0.011$ & $1.430\pm0.006$ & $0.111\pm0.004$\\
\hline
\end{tabular}
\end{table}

\end{landscape}
}

\begin{figure*}
	\includegraphics[width=\textwidth]{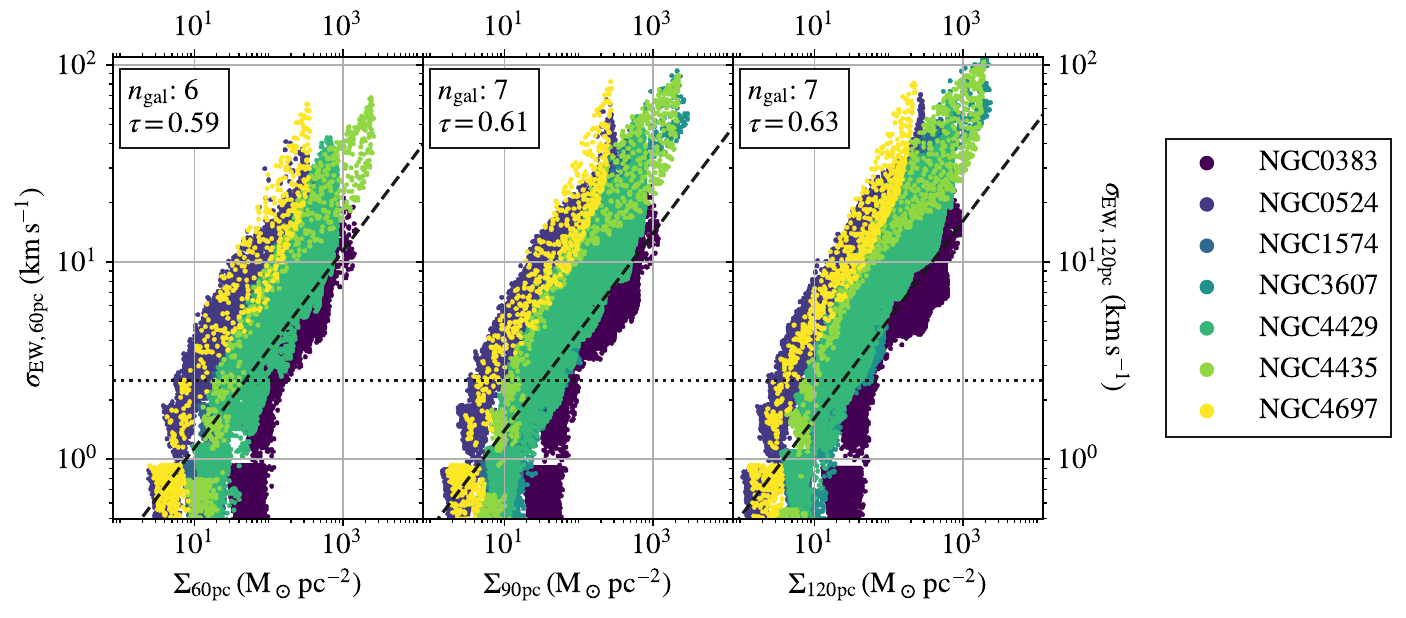}
    \caption{$\sigma_\mathrm{EW}$ -- $\Sigma$ relations of our seven WISDOM sample galaxies, at a number of spatial resolutions. In each panel, the colour of the data points corresponds to the galaxy (coloured in numerical order; legend on the right-hand side). The black dashed line shows the fiducial virial equilibrium relation. The black horizontal dotted line is the channel width of our data. The number of galaxies included in the particular panel and the Kendall-$\tau$ correlation coefficient of the data points are listed in the top-left corner.}
    \label{fig:sigma_sigma}
\end{figure*}

The beam-by-beam $\sigma_\mathrm{EW}$ -- $\Sigma$ relation for the entire sample is shown in Fig.~\ref{fig:sigma_sigma}. In general, the majority of the data points of each galaxy are above the fiducial virial equilibrium line. There is also a significant scatter, of about $0.3$~dex in effective line width at any given $\Sigma$, which is predominantly driven by galaxy-to-galaxy variations (the scatter within any individual galaxy is only $\approx0.2$~dex\footnote{This scatter reflects both the intrinsic scatter $\Delta$ and the scatter due to measurement uncertainties.}, versus $0.3$~dex across the full sample). Compared to the best-fitting parameters in \cite{2018Sun}, $\beta$ and $A$ are slightly larger here. The power-law index $\beta\gtrsim0.6$ in ETGs compared to $\approx0.5$ in star-forming galaxies\footnote{This coefficient is not reported in either \citet{2018Sun} or \citet{2020Sun} for the centres of their barred galaxies, but a visual inspection of figure~2 in \citet{2020Sun} suggests a similar $\beta$ in this regime.}, although this difference is within the intrinsic scatters of the fits. The zero-point $A\approx1$ in ETGs compared to $\approx0.85$ in star-forming galaxies, implying velocity dispersions that are $\approx40\%$ higher at a fixed molecular gas mass surface density in ETGs than in star-forming galaxies. The intrinsic scatters about the relations $\Delta$ are however similar (typically $\approx0.1$). These indicate relations that are generally steeper in ETGs than in star-forming galaxies, and a clear offset above the fiducial virial equilibrium line.

Whilst we adopt the formalism above to interpret our effective line width measurements, we recall that it assumes that clouds are isolated spherical entities, and that our beam-by-beam measurements effectively probe cloud-scale structures. These constitute a useful limiting case, but are a crude description of reality. An alternative useful case to consider is that of a thin disc in vertical dynamical equilibrium, where the relation between mass surface density and vertical velocity dispersion is
\begin{equation}
    \sigma_\mathrm{RMS,z}(R)=\sqrt{\pi cG\Sigma h_\mathrm{z}}
\end{equation}
\citep[e.g.][]{1988VanDerKruit}, where $c$ varies between $3/2$ (exponential disc) and $2$ (isothermal disc) and $h_\mathrm{z}$ is the vertical scale height. For an order-of-magnitude calculation, we take $c=3/2$ and $h_\mathrm{z}=100$~pc, as in e.g.\ the MW thin disc \citep[e.g.][]{2008Juric, 2008KongZhu}. In that case, for $\Sigma=100$~$\mathrm{M_\odot~pc^{-2}}$ one obtains $\sigma_\mathrm{RMS,z}\approx10$~km~s$^{-1}$, similar to the effective line widths measured across our sample. This provides confidence that our assumption of the synthesised beam size being the relevant scale is reasonable.

\subsection{Virial parameter}\label{sec:virial_parameter}

Using our effective line width and molecular gas mass surface density measurements, we can estimate a quantity analogous to the virial parameter $\avir$, that quantifies the balance between kinetic energy and gravity. For a nearly spherical cloud,
\begin{equation}\label{eq:virial_eq}
\alpha_\mathrm{vir}\equiv\frac{5\,\sigma_\mathrm{RMS}^2R}{f\,GM}
\end{equation}
\citep[e.g.][]{1992BertoldiMcKee, 2018Sun}, where $f$ is a geometrical factor depending on the cloud radial mass volume density profile $\rho(r)$. For $\rho(r)\propto r^{-\gamma}$, where $r$ is the radius and $\gamma$ the power-law index, $f=(1-\gamma/3)/(1-2\gamma/5)$. Following e.g.\ \citet{2006RosolowskyLeroy} and \citet{2018Sun}, we assume the gas structures in our sample galaxies have a radial density profile $\rho(r)\propto r^{-1}$ and thus $f=10/9$. This fixed $\gamma$ assumption reflects the fact that we expect a similar sub-beam cloud density profile across our galaxies, while the specific $\gamma$ adopted allows  immediate comparison to earlier works \citep[e.g.][]{2006RosolowskyLeroy, 2018Sun}. This approach is highly idealised, as the interstellar medium (ISM) has a complex structure, but it at least allows to investigate relative $\avir$ differences and compare with other works.

Using the effective line width and molecular gas mass surface density measurements from our beam-by-beam analysis, we can construct and estimate a quantity $\abeamvir$ analogous to the virial parameter \citep{2018Sun}:
\begin{equation}\label{eq:virial_eq_beam}
    \abeamvir\equiv5.77\,\left(\frac{\sigma_\mathrm{EW}}{\mathrm{km~s^{-1}}}\right)^2\,\left(\frac{\Sigma}{\mathrm{M_\odot~pc^{-2}}}\right)^{-1}\,\left(\frac{r_\mathrm{beam}}{\mathrm{40~pc}}\right)^{-1},
\end{equation}
where $r_\mathrm{beam}$ is simply the spatial scale (i.e.\ the spatial resolution) of the map in question divided by $2$, as this is a radius rather than a diameter. This expression effectively assumes that the size of the synthesised beam is equal to the cloud size, which we expect to be approximately true (to within a factor of a few) given the spatial scales considered here. However, for small clouds this will underestimate $\abeamvir$, conversely for large clouds. We calculate this modified virial parameter at each of the three spatial scales considered for each of our sample galaxies, and show an example of the distribution (weighted by $\Sigma$) for NGC~4429 in Fig.~\ref{fig:virial_parameter_ngc4429}. Analogous figures for the other sample galaxies are shown in Appendix~\ref{app:avir}. The median $\abeamvir$ is elevated, larger than $\abeamvir=1$ expected from virial equilibrium, and in fact the majority of sight lines are significantly above the marginal gravitational boundness limit $\abeamvir=2$. This suggests that either the gas must be confined by some external non-gravitational force, or the gas structures probed must be short lived \citep[see e.g.][]{2021Liu}. The $\abeamvir$ distribution remains relatively constant as a function of spatial scale, however, suggesting that the gas follows a constant `scale -- linewidth' relation (this is also clear in the similar relation of Fig.~\ref{fig:sigma_sigma_ngc4429}).

\begin{figure*}
    \includegraphics[width=\textwidth]{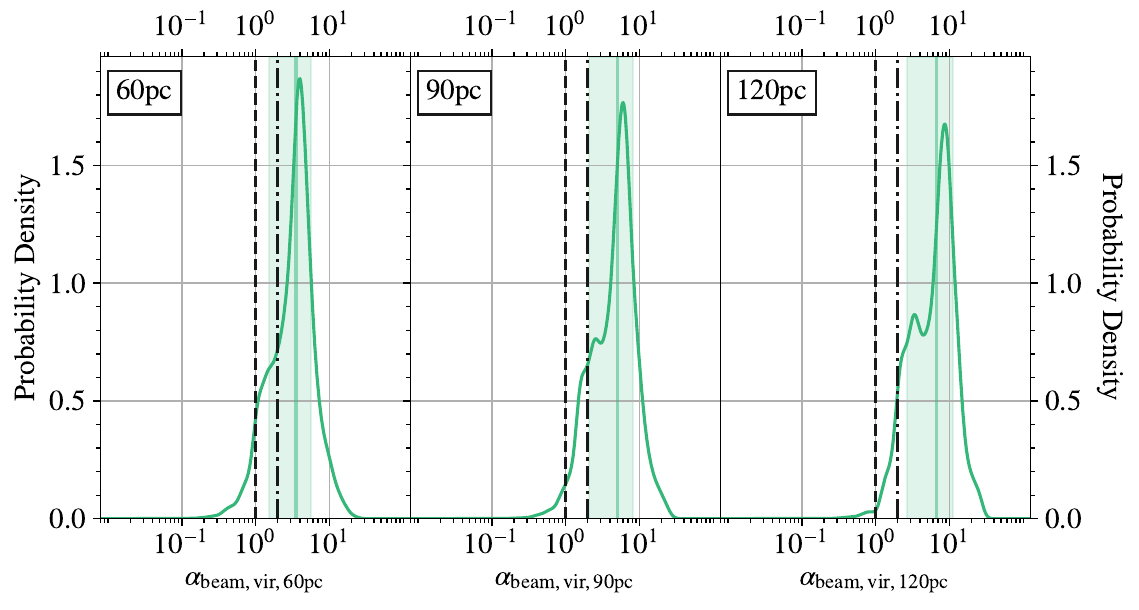}
    \caption{Distributions of $\abeamvir$ for all the spaxels of NGC~4429. In each panel (i.e.\ each spatial scale), we indicate the median of the distribution as a green solid line, while the green shaded region indicates the $16^\mathrm{th}$ to $84^\mathrm{th}$ percentile range. The fiducial virial equilibrium ($\abeamvir=1$) is indicated by the black dashed line and the fiducial marginal gravitational boundness ($\abeamvir=2$) by the black dot-dashed line. The spatial scale is listed in the top-left corner.}
    \label{fig:virial_parameter_ngc4429}
\end{figure*}

\begin{figure}
	\includegraphics[width=\columnwidth]{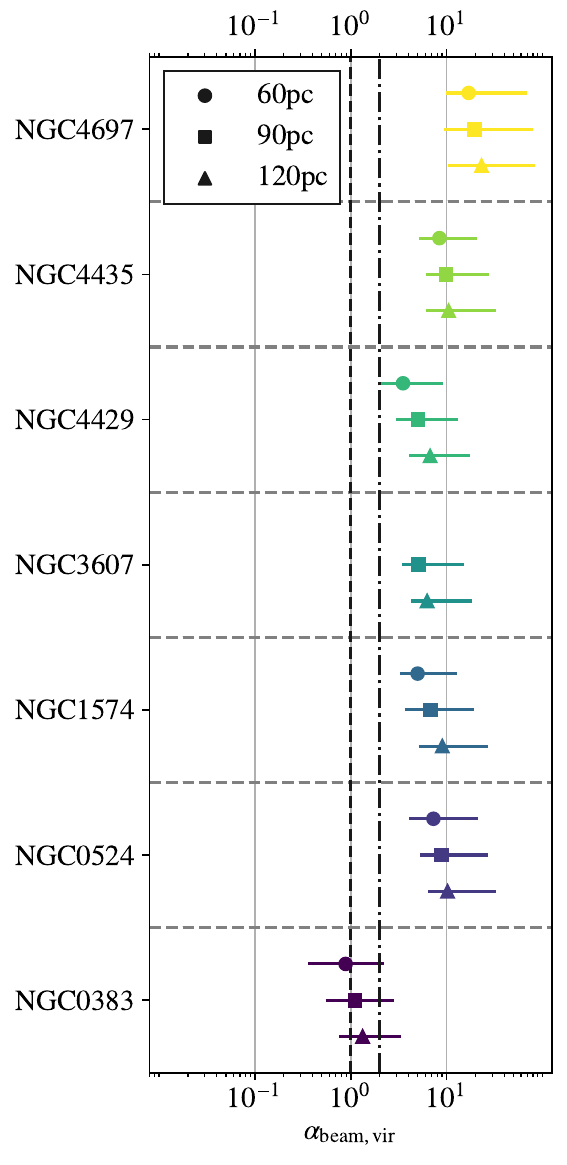}
    \caption{Modified virial parameter $\abeamvir$ for all the WISDOM sample galaxies, at a number of spatial scales (i.e.\ spatial resolutions). Galaxies are ordered alphabetically from bottom to top. In each case, the median of the distribution is shown by the solid symbol, and the $16^\mathrm{th}$ to $84^\mathrm{th}$ percentile range is indicated by the horizontal line. The black dashed line indicates fiducial virial equilibrium ($\abeamvir=1$) and the black dot-dashed line fiducial marginal gravitational boundness ($\abeamvir=2$).}
    \label{fig:virial_par_overview}
\end{figure}

We show the $\abeamvir$ distributions of all our sample galaxies in Fig.~\ref{fig:virial_par_overview}. With the exception of NGC~0383, similar trends are present across all sample galaxies, i.e.\ $\abeamvir$ is generally higher than $1$. At a spatial scale of $60$~pc, the molecular gas mass surface density-weighted median is $\abeamvir=4.2^{+7.3}_{-3.2}$, where the uncertainties represent the $16^{\rm th}$ and $84^{\rm th}$ percentiles of the distribution, weighting each galaxy equally (to account for the different number of synthesised beams across each galaxy). These modified virial parameters are similar to those measured in the centres of barred galaxies by \cite{2020Sun}, where the median $\abeamvir$ of their sample galaxies is $6$. Although there are exceptions, as shown here by the case of NGC~0383 \citep[and previously by NGC~4526;][]{2015Utomo}, in general the gas in ETGs is unlikely to be in virial equilibrium. A similar result was obtained by \citet{2021Liu}, who used a more traditional cloud extraction method and showed that the (effective) cloud virial parameters are generally larger than $1$ in NGC~4429 ($\langle\alpha_\mathrm{eff,vir}\rangle\approx2.2$).

We note that galactic rotation can impact the measured effective line widths, due to beam smearing of unresolved velocity gradients (that becomes more important at worse spatial resolutions). However, the distributions shown in Fig.~\ref{fig:virial_par_overview} only have a weak scale dependence, suggesting that this effect is unimportant in our high-spatial resolution data (that typically spatially resolve clouds). Nevertheless, for observations with worse spatial resolutions, the impact of galactic rotation on the measured effective line widths may well need to first be subtracted. 

\subsection{Pressure estimates}\label{sec:pressure}

The balance between internal and external pressure in the molecular gas is an important diagnostic of the dynamical state of the ISM. Here, we calculate these quantities across our sample galaxies on a beam-by-beam basis, and again investigate whether the molecular gas is in dynamical equilibrium.

\subsubsection{Internal turbulent pressure}

Using our beam-by-beam molecular gas measurements, we estimate the internal turbulent pressure $\pturb$, or equivalently the kinetic energy volume density, of the molecular gas as
\begin{equation}\label{eq:p_turb}
    \frac{\pbeamturb/k_\mathrm{B}}{\mathrm{K~cm^{-3}}}\approx61.3\,\left(\frac{\Sigma}{\mathrm{M_\odot~pc^{-2}}}\right)\,\left(\frac{\sigma_\mathrm{EW}}{\mathrm{km~s^{-1}}}\right)^2\,\left(\frac{r_\mathrm{beam}}{\mathrm{40~pc}}\right)^{-1}
\end{equation}
\citep{2018Sun}. As can be seen, $\pbeamturb\propto\sigma_\mathrm{EW}^2\Sigma$ whilst $\abeamvir\propto\sigma_\mathrm{EW}^2\Sigma^{-1}$ (Eq.~\ref{eq:virial_eq_beam}), so while these two quantities are somewhat related, they are not completely degenerate. We show an example of the $\pbeamturb$ distribution (weighted by $\Sigma$) for NGC~4429 in Fig.~\ref{fig:pturb_ngc4429}. Analogous figures for the other sample galaxies are shown in Appendix~\ref{app:pturb}. The $\pbeamturb$ distributions of all our sample galaxies are shown in Fig.~\ref{fig:turb_pressure_overview}. As for $\abeamvir$, there are large variations both within and across galaxies. The turbulent pressures can span an order of magnitude or more within a single galaxy, indicating that they are a strong function of local galactic conditions, and they vary by about three orders of magnitude across the whole sample, showing an additional dependence on global galactic conditions. At a spatial scale of $60$~pc, and weighting each galaxy equally, the median molecular gas mass surface density-weighted $\pbeamturb$ is $1.1^{+24.2}_{-1.0}\times10^7$~K~cm$^{-3}$, where the uncertainties again represent the $16^{\rm th}$ and $84^{\rm th}$ percentiles of the distribution. And again, as for $\abeamvir$, the $\pbeamturb$ distributions show evidence of a scale dependence. We note that these $\pbeamturb$ are significantly larger than those of the discs of star-forming galaxies, that are typically $\approx2\times10^4$~K~cm$^{-3}$ \citep{2020Sun}. Without weighting by mass surface density or galaxy, the median $\pbeamturb$ of our sample galaxies is $3.5\times10^6$~K~cm$^{-3}$, similar to that of the centres of barred galaxies \citep[$5.1\times10^6$~K~cm$^{-3}$;][]{2020Sun}  and in some cases approaching the turbulent pressures of MW central molecular zone (CMZ) clouds \citep[$10^9$~K~cm$^{-3}$; e.g.][]{1988Bally}. This overpressured environment has been postulated to be the reason for the low star-formation efficiency of the MW CMZ \citep{2014Kruijssen}, and we discuss this further in Section~\ref{sec:cmz_discussion}. Simulations have also found very high turbulent pressures in the centres of bulge-dominated galaxies \citep{2020Gensior}, and our results appear to confirm this finding.

\begin{figure*}
	\includegraphics[width=\textwidth]{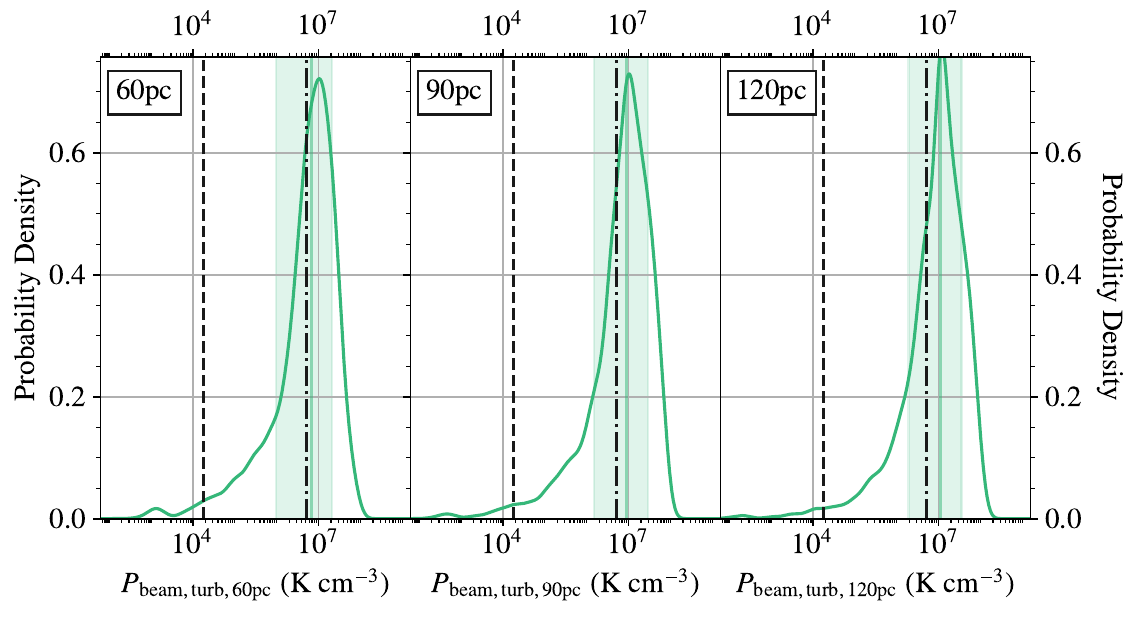}
    \caption{Distributions of $\pbeamturb$ for all the spaxels of NGC~4429. In each panel (i.e.\ each spatial scale), we indicate the median of the distribution as a green solid line, while the green shaded region indicates the $16^\mathrm{th}$ to $84^\mathrm{th}$ percentile range. The black dashed line indicates the average $\pbeamturb$ across the discs of LTGs, while the black dot-dashed line indicates that in the centres of barred galaxies, both from \citet{2020Sun}. The spatial scale is listed in the top-left corner.}
    \label{fig:pturb_ngc4429}
\end{figure*}

\begin{figure}
	\includegraphics[width=\columnwidth]{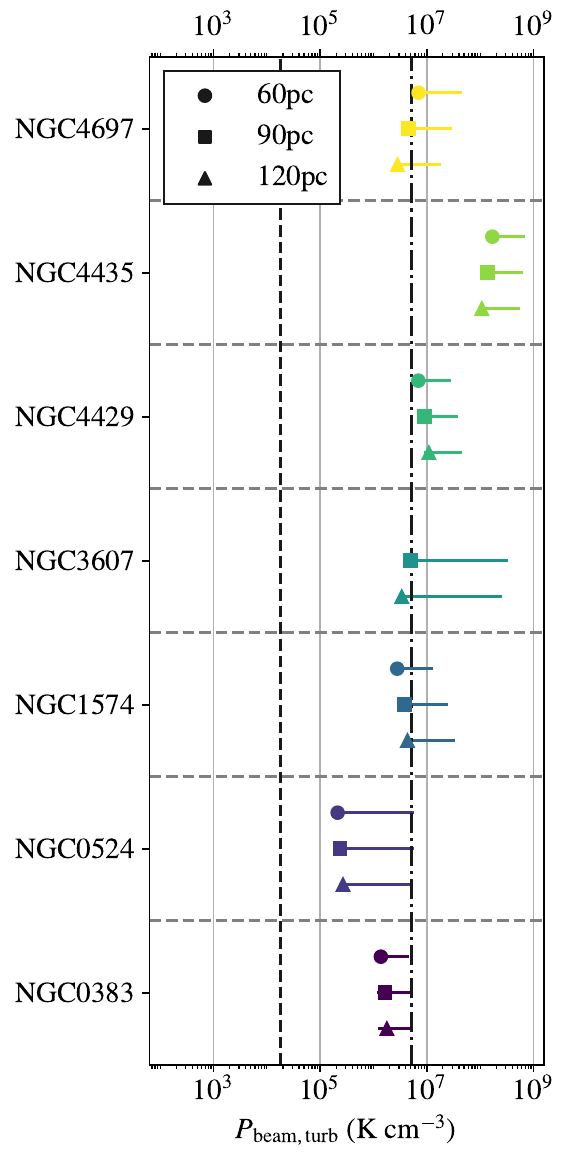}
    \caption{Turbulent pressure $\pbeamturb$ for all the WISDOM sample galaxies, at a number of spatial scales (i.e.\ spatial resolutions). Galaxies are ordered alphabetically from bottom to top. In each case, the median of the distribution is shown by the solid symbol, and the $16^\mathrm{th}$ to $84^\mathrm{th}$ percentile range is indicated by the horizontal line. The black dashed line indicates the average $\pbeamturb$ measured at $150$~pc scale across the disc of LTGs, while the black dot-dash indicates that in the centres of barred galaxies, both from \citet{2020Sun}.}
    \label{fig:turb_pressure_overview}
\end{figure}

\subsubsection{Dynamical equilibrium pressure}

The dynamical equilibrium pressure, $\pde$, is the pressure required for the molecular gas to remain in vertical equilibrium with the gravitational potential of the galaxy. It therefore requires knowledge of both the gas mass and the stellar mass:
\begin{equation}\label{eq:p_de}
    \pde=\frac{\pi G}{2}\Sigma_\mathrm{gas}^2+\sqrt{2G\rho_\ast}\,\Sigma_\mathrm{gas}\sigma_\mathrm{gas}\,,
\end{equation}
\citep[][their eq.~12]{2020aSun}, where $\rho_\ast$ is the stellar mass volume density and quantities with the subscript `gas' indicate contributions from both the molecular and atomic gas. We note that this relationship assumes that the gas and stars each lie in an isothermal discs (i.e.\ that they each have a $\mathrm{sech}^2$ vertical mass volume density profile) and that the stellar scale height is much larger than that of the gas, although the former assumption may not be valid for large spheroids. As we do not have sub-arcsecond resolution H{\sc i} data with which to measure the mass surface density and velocity dispersion of the atomic gas, and as we expect the molecular gas to dominate in the central regions of our sample ETGs \citep[e.g.][]{2010SerraOosterloo, 2023Maccagni}, we neglect the atomic gas terms and $\Sigma_\mathrm{gas}$ and $\sigma_\mathrm{gas}$ simply become $\Sigma$ and $\sigma_\mathrm{RMS}$, respectively. We note that any contribution from atomic gas will make $\Sigma_\mathrm{gas}$ larger, and since the typical velocity dispersion of H{\sc i} gas is $\approx10$~km~s$^{-1}$ \citep{2008Leroy}, this approximate $\pde$ should formally be considered a lower limit (although this may be more complex in galaxy centres, where the velocity dispersions are higher).

To calculate the stellar mass volume densities, we use the available MGE stellar surface brightness models of our sample galaxies (see Section~\ref{sec:data} and Table~\ref{tab:data_overview}),
that can trivially be converted into mass surface density models using the mass-to-light ratios provided and then into mass volume density models through an analytic deprojection \citep[see eqs.~1 and 6 of][assuming $p=1$ for axisymmetry]{2002Cappellari}. This allows to calculate the stellar volume densities at the same spatial resolutions as our ALMA data, and in turn to calculate the $\pbeamturb/\pde$ ratios in a spatially-resolved manner, where we again assume $\sigma_\mathrm{RMS}\approx\sigma_\mathrm{EW}$.

\begin{figure*}
	\includegraphics[width=\textwidth]{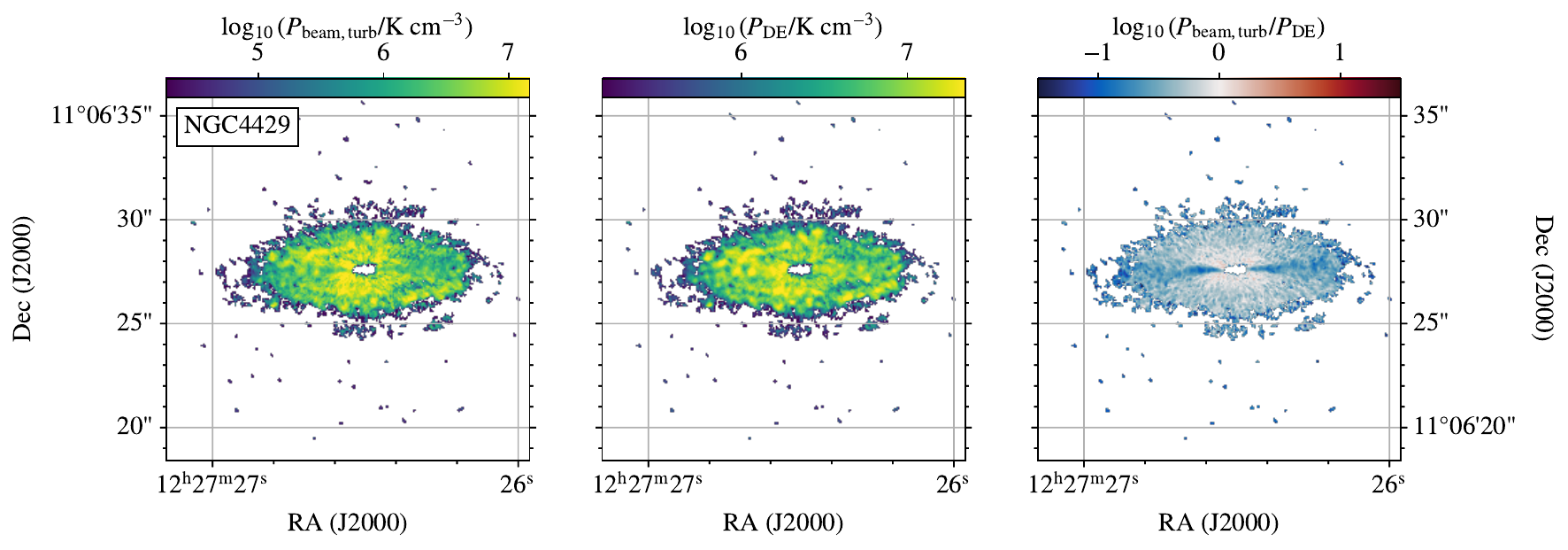}
    \caption{Maps of the turbulent pressure $\pbeamturb$ (left; measured at the native spatial resolution), the dynamical equilibrium pressure $\pde$ (middle) and the pressure ratio $\pbeamturb/\pde$ (right). For the two pressure maps, the colour bar ranges from the $2.5^\mathrm{th}$ to the $97.5^\mathrm{th}$ percentile of the (logarithmic) distribution. For the pressure ratio map, the colour bar is scaled such that a ratio of $1$ is white, and increasingly high $\pbeamturb$ ($\pde$) appear as red (blue). The biconical shape along the galaxy minor axis is due to beam smearing, and as expected is also clearly present in the velocity dispersion map (see Fig.~\ref{fig:data}).}
    \label{fig:pressure_ratio_ngc4429}
\end{figure*}

We show an example of a $\pbeamturb/\pde$ pressure ratio map for NGC~4429 in Figure~\ref{fig:pressure_ratio_ngc4429}. The dynamical equilibrium pressure $\pde$ is significantly larger than the the turbulent pressure $\pbeamturb$ across the entire molecular gas disc, i.e.\ $\pbeamturb/\pde\le1$ throughout. This indicates that the gas is not in dynamical equilibrium, and at first glance appears to show that the gas is strongly gravitationally bound, so star formation should be widespread \citep{2022OstrikerKim}. This result is also unexpected and counter-intuitive, given that our $\abeamvir$ measurements indicate the gas should be unbound ($\abeamvir>1$). However, it is important to note that $\pbeamturb$ is a three-dimensional quantity, whilst $\pde$ acts purely vertically (i.e.\ into the mid-plane). As such, it may be that clouds are strongly bound in the $z$-direction but are (more) unbound in the plane (radially and/or azimuthally), which would lead to cloud elongation in the plane (e.g.\ along the radial direction, as observed by \citealt{2021Liu} in NGC~4429). Indeed, although not considered in this work, shear and tides could be important, and would increase the timescales for collapse and lead to less gravitationally bound gas. As mentioned above, we also note that the $\pde$ formalism (Eq.~\ref{eq:p_de}) assumes that the stars are in an isothermal disc, that is unlikely to be the case in the central regions of our sample ETGs. The true stellar distributions may well be more vertically extended, which would likely lower $\pde$ somewhat. Exploring the potential effects of the large-scale mass distributions and dynamics is beyond the scope of this work, however, and we leave a full exploration of these effects to a future study. Analogous figures for the other sample galaxies are shown in Appendix~\ref{app:pressure_ratio}.

\begin{figure*}
	\includegraphics[width=\textwidth]{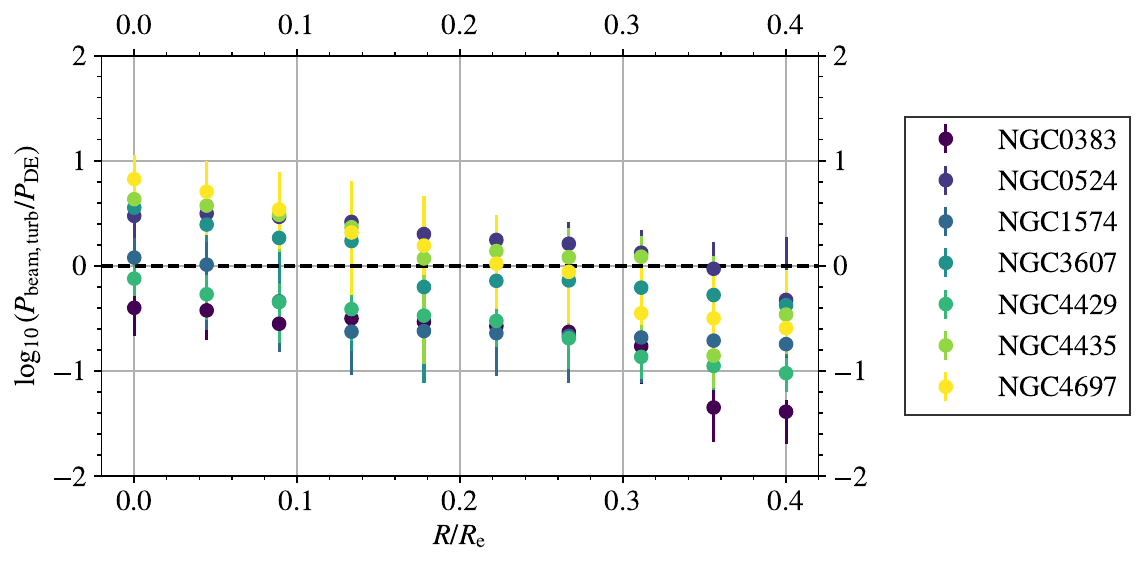}
    \caption{Radial profiles (measured at the native resolution and averaged azimuthally) of the ratio of internal turbulent pressure to dynamical equilibrium pressure $\pbeamturb/\pde$ for all the WISDOM sample galaxies. The radii are normalised by the effective radius $R_\mathrm{e}$ of each galaxy. The black horizontal dashed line indicates $\pbeamturb/\pde=1$.}
    \label{fig:pressure_comparison_radial}
\end{figure*}

We calculate azimuthally-averaged radial profiles of the $\pbeamturb/\pde$ ratios of all our sample galaxies, using elliptical annuli based on the position angles and inclination angles reported in the references of Table~\ref{tab:data_overview}. These profiles are shown in Figure~\ref{fig:pressure_comparison_radial}, where the radii have been normalised by the effective radius of each galaxy. The majority of the galaxies again have $\pde>\pbeamturb$ across most of their molecular gas disc, the exception being the galaxy centres (at least partially due to beam smearing and unresolved velocity structures). The clear outlier is NGC~0524, that has $\pbeamturb>\pde$ almost everywhere. NGC~0383 is the one galaxy that behaves as expected in term of both its $\abeamvir$ and $\pbeamturb/\pde$ parameters: it has the lowest $\pbeamturb/\pde$ ratio, along with the highest SFR and lowest $\abeamvir$, as expected for strongly gravitationally bound gas. Whilst \cite{2020aSun} showed that for star-forming galaxies the gas is approximately in pressure equilibrium (i.e.\ the pressure ratios are about unity, with slight increases in the galaxy centres), this is clearly not the case in our ETGs. The most discrepant galaxy in this case is NGC~4697, that has extremely low velocity dispersions \citep[$\approx1.5$~km~s$^{-1}$;][]{2017Davis}, thus the estimated $\pde$ may be (significantly) overestimated (as our velocity dispersion measurements are likely to be biased high).

\section{Caveats and Discussion}\label{sec:discussion}

\subsection{Potential variation in the CO-conversion factor}\label{sec:alpha_co_variation}

Throughout this work, we have used a canonical $\aco$ suitable for the MW, which is often employed in the literature \citep[e.g.][]{2018Sun, 2021Liu, 2022Maeda}. We refer readers to the review by \cite{2013Bolatto} for a comprehensive discussion of $\aco$ variations within galaxies. In this context, it is important to note that a number of works find $\aco$ variations in galaxy centres, where it is often determined to be lower than the standard MW $\aco$ \citep{2013Sandstrom,2022Teng}. The precise reason for this is unknown, but given that galaxy centres are particularly dynamically active environments, this is not entirely surprising. The picture is also completely unclear for ETGs, currently lacking any measurement of $\aco$ (without assuming that clouds are in virial equilibria). Clearly, more work is needed to constrain the conversion factor in galaxy centres, either through multi-line modelling \citep{2022Teng} or through simultaneously constraining the dust-to-gas ratio and $\aco$ \citep{2013Sandstrom}. 

If we assume the gas in ETGs to be similar to that in the CMZs of LTGs, with an approximately constant $\aco$ over the central kiloparsec (their typically extent), then if $\aco$ were to be lower \citep[as found in some spiral galaxies; e.g.][]{2013Sandstrom,2022Teng} our results would become even more discrepant -- $\abeamvir$ would increase (see Eq.~\ref{eq:virial_eq_beam}) and  $\pbeamturb$ would decrease (see Eq.~\ref{eq:p_turb}). The gas would then be \emph{further} out of dynamical equilibrium. We would also then require $\aco$ to be around an order of magnitude \emph{larger} to bring our results in line with those of LTGs; this seems unlikely, so we believe the general trend of our results to be robust.

\subsection{Comparison to LTGs}\label{sec:comparison_ltgs}

\begin{figure*}
	\includegraphics[width=\textwidth]{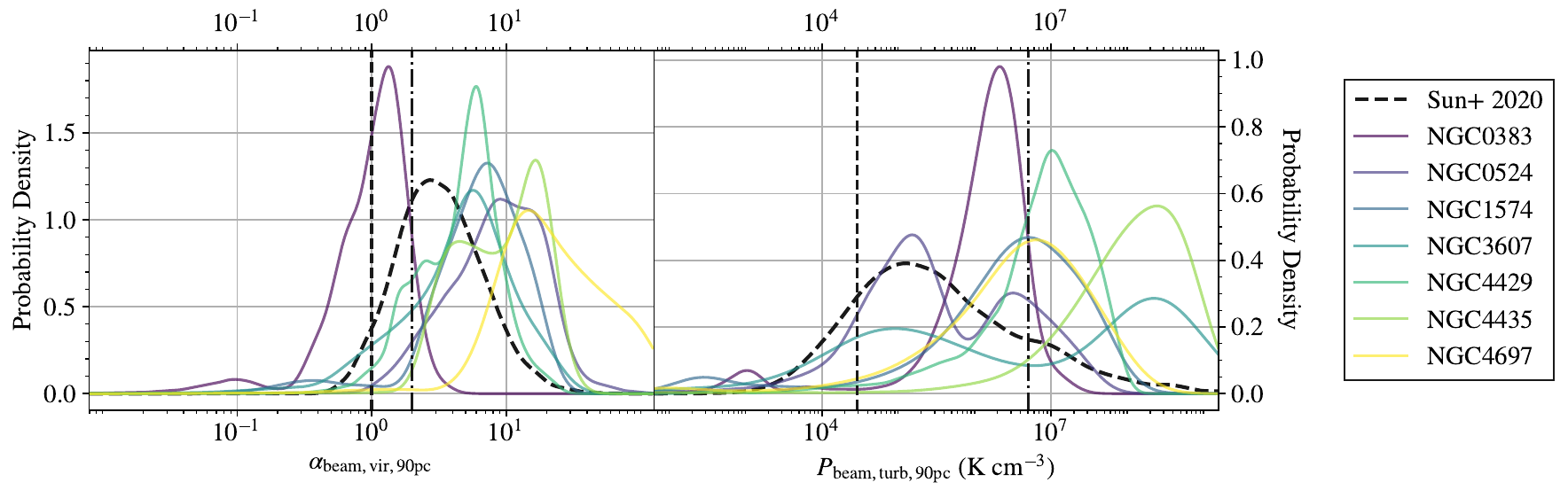}
    \caption{Comparison of the $\abeamvir$ and $\pbeamturb$ distributions of ETGs to those of the LTG sample of \citet{2020Sun}, as measured across the entire discs of their star-forming sample galaxies. For each galaxy, we calculate a molecular gas mass surface density-weighted kernel density estimator (KDE) at a common resolution of $90$~pc. For $\abeamvir$, the black dashed line indicates fiducial virial equilibrium ($\abeamvir=1$) and the black dot-dashed line fiducial marginal gravitational boundness ($\abeamvir=2$). For $\pbeamturb$, the black dashed line indicates the average $\pbeamturb$ across the discs of LTGs (measured at $90$~pc resolution), while the black dot-dashed line indicates that in the centres of barred galaxies (measured at $150$~pc resolution), both from \citet{2020Sun}.}
    \label{fig:sun_comparison}
\end{figure*}

ETGs offer a very different perspective of the gas than the more widely studied star-forming spiral galaxy population. The molecular gas of ETGs is typically concentrated in the central kiloparsec. It is also typically much smoother and is a strong function of morphology \citep{2022Davis}. As such, it is informative to compare its properties to those of the molecular gas of regular star-forming galaxies. In Figure~\ref{fig:sun_comparison}, we compare the $\abeamvir$ and $\pbeamturb$ distributions of the PHANGS sample \citep{2020Sun}, calculated on a beam-by-beam basis, to those of the ETG galaxies in this work (calculated in an analogous manner). We use a common resolution of $90$~pc, where our chosen resolutions overlap with that of the \cite{2020Sun} sample. The differences are striking -- the gas of LTGs typically has $\abeamvir\approx2$, whereas for ETGs (with the exception of NGC~0383) $\abeamvir$ is significantly higher. The cause of the low $\abeamvir$ of NGC~0383 is puzzling -- it has the highest SFR of our sample galaxies, and is also the only galaxy in our sample with radio jets \citep[e.g.][]{1968Macdonald}. Given our small sample size, binning by radio activity or SFR is not possible, but clearly there is some diversity in the gas properties of ETGs. The ETG $\abeamvir$ are even higher than those of the most extreme centres of barred galaxies, where $\abeamvir\approx6$ \citep{2020Sun}.

\begin{figure*}
	\includegraphics[width=\textwidth]{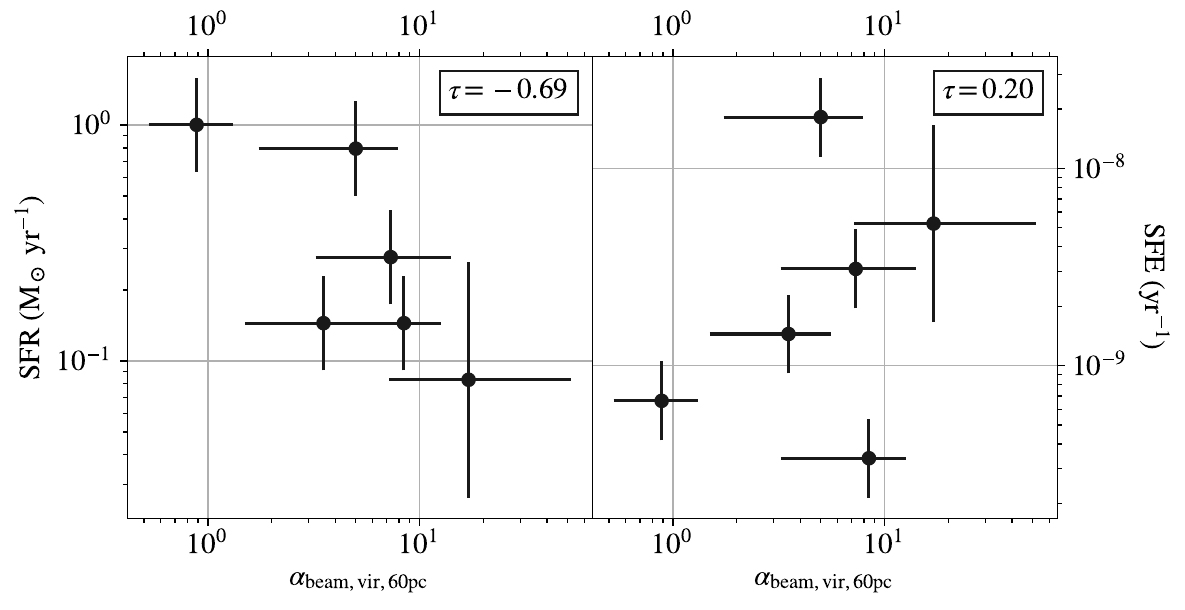}
    \caption{{\it Left}: SFR versus $\abeamvir$ relation for our sample galaxies, using the average $\abeamvir$ measurements at a resolution of $60$~pc. {\it Right}: Analogous SFE versus $\abeamvir$ relation. The Kendall-$\tau$ correlation coefficient of the data points is indicated in the top-right corner of each panel. Analogous plots for other resolutions are shown in Appendix \ref{app:avir_sfr}.}
    \label{fig:alpha_vir_sfr}
\end{figure*}

LTGs also typically have turbulent pressures lower than those of ETGs, by an order of magnitude or more. The conclusions reached by \cite{2020Sun} and Hughes et al.\ (in preparation) are that the molecular clouds of star-forming galaxies are gravitationally bound, long-lived features of the galactic discs. Our results however indicate the opposite in ETGs: the gas of ETGs is likely not bound. This suggests that the gas may remain concentrated into `clouds' for less time than the $\sim10$~Myr required for star-formation to initiate \citep[e.g.][]{2022Kim}. To highlight this, the left panel of Figure~\ref{fig:alpha_vir_sfr} shows the correlation of the average $\abeamvir$ and the global SFRs of our sample galaxies. Despite the small sample size (in this case, six galaxies), there tentatively appears to be an anti-correlation between $\abeamvir$ and SFR. Very similar results are obtained at other resolutions (Appendix \ref{app:avir_sfr}). This difference of the gas conditions offers a natural explanation for the suppression of star formation in ETGs relative to LTGs \citep[e.g.][]{2014Davis}. However, SFE may be the more fundamental parameter here, and we show the analogous correlation of the average $\abeamvir$ and the golabl SFEs of our sample galaxies in the right panel of Figure \ref{fig:alpha_vir_sfr}, where we have used the total molecular gas masses of \cite{2022Davis} to calculate the global (molecular gas) SFEs. The trend is weaker (with a lower $\tau$) than and in the opposite direction to that with SFR. This may indicate that the virial parameters are not particularly strongly related to the star formation activity of these ETGs, but to draw stronger conclusions would require a larger sample.

\subsection{Is the molecular gas of ETGs similar to that of CMZs?}\label{sec:cmz_discussion}

\begin{figure*}
	\includegraphics[width=\textwidth]{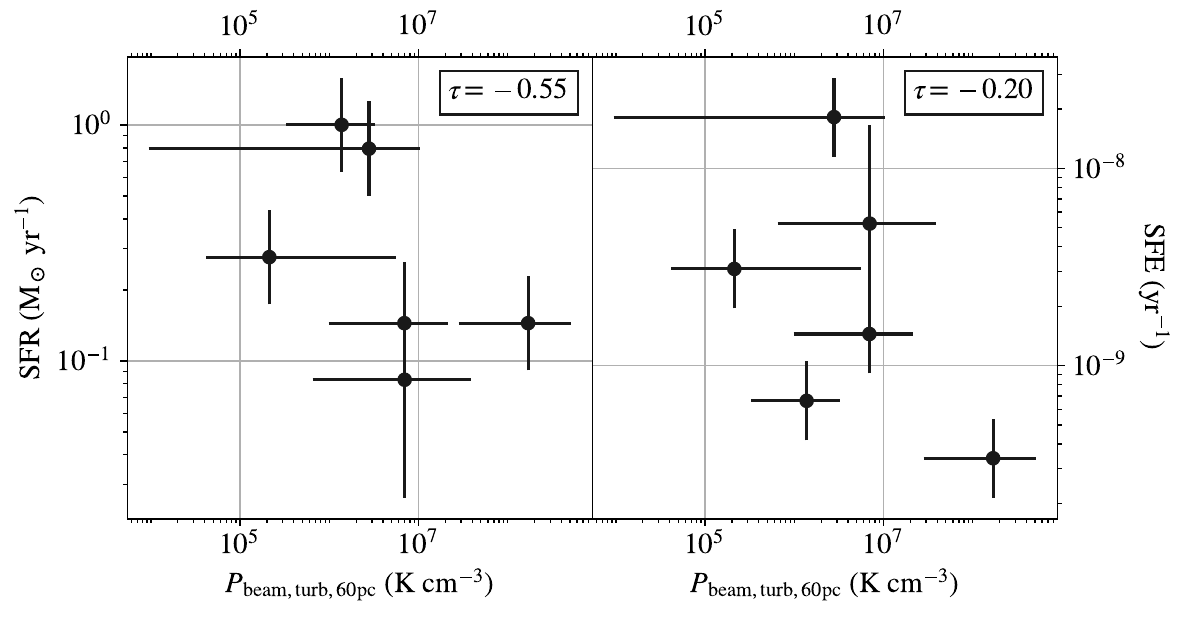}
    \caption{{\it Left}: SFR versus $\pbeamturb$ relation for our sample galaxies, using the average $\pbeamturb$ measurements at a resolution of $60$~pc. {\it Right}: Analogous SFE versus $\pbeamturb$ relation. The Kendall-$\tau$ correlation coefficient of the data points is indicated in the top-right corner of each panel. Analogous plots for other resolutions are shown in Appendix \ref{app:pturb_sfr}.}
    \label{fig:pressure_sfr}
\end{figure*}

At least in terms of turbulent pressure, the gas in ETGs seems to be similar to that of the CMZ of the MW \citep[e.g.][]{1988Bally}. Much like our sample ETGs, the CMZ has suppressed star formation relative to the rest of the MW disc, and it has been proposed that the turbulent pressure of the gas could be the culprit \citep{2014Kruijssen}. \cite{2005KrumholzMcKee} suggest that star-formation occurs only in regions where the gravitational potential exceeds the internal turbulent motions, and so for more turbulent regions the effective critical density for star formation to occur increases. For our sample galaxies, we show the relation between the average $\pbeamturb$ and the global SFR in the left panel of Figure~\ref{fig:pressure_sfr}, revealing a tentative anti-correlation (much like $\abeamvir$ and SFR in Fig.~\ref{fig:alpha_vir_sfr}). The right panel of Figure~\ref{fig:pressure_sfr} shows the analogous trend with global (molecular gas) SFE, that is in the same direction but weaker (lower $\tau$). This trend also holds at other resolutions, albeit with somewhat weaker correlation coefficients (Appendix \ref{app:pturb_sfr}). High pressures in the centres of ETGs seem to be relatively ubiquitous, so this may be a general result for this class of galaxies. Although in the very centres of galaxies this could be due to beam smearing, these high pressures are prevalent across the entirety of the CO discs. Given our high spatial resolutions, beam smearing is not expected to be an issue across the entire CO extents, so this result is likely to be physical rather than an observational bias. The turbulent pressures are also higher in the centres of barred galaxies, as shown by \citet{2020Sun} and highlighted in Fig.~\ref{fig:sun_comparison}.

Our results also show that the external gravitational potential is a significant factor regulating the cloud lifecycle, which \citet{2020Meidt} showed may naturally explain the low SFE of the MW CMZ. However, not all star-forming galaxies have a suppression of star formation in the centre, and many in fact show enhanced central star formation \citep{2021Querejeta}, although this may depend on the particular gas tracer used \citep{2015Usero}. Further work in this direction should focus on improving the spatial resolution of the observations to cleanly isolate the CMZs of external galaxies, and {\it JWST} is ideal for high-resolution mapping of the stellar and SFR distributions in galaxy centres, as increased dust extinction often complicates this in the optical regime. One may also obtain significantly different results when using denser gas tracers, as dense gas is more closely associated with star formation \citep[e.g.][]{2004GaoSolomon}. Focusing on HCN (or other dense gas tracers) rather than CO should thus offer important insights into the suppression or enhancement of star formation in these more extreme galactic environments.

\section{Conclusions}\label{sec:conclusions}

In this work, we have presented an overview of the beam-by-beam properties of the molecular gas in seven early-type galaxies (ETGs). These quiescent galaxies have been shown to have star formation efficiencies \citep[SFEs][]{2014Davis} and morphologies \citep{2022Davis} that are significantly different from those of the star-forming galaxies more widely studied in the literature \citep[e.g.][]{2010Hughes, 2020Sun, 2021Rosolowsky}. This work now shows that the properties of the molecular gas in these ETGs are also very different.
% from the those of the spiral galaxies discussed in the literature.
Our main results are as follows:
\begin{enumerate}
    \item The molecular gas of individual ETGs typically spans two orders of magnitude in molecular gas mass surface density $\Sigma$ ($10$ -- $1000$~$\mathrm{M_\odot~pc^{-2}}$) and one order of magnitude in velocity dispersion $\sigma_\mathrm{EW}$ (from our velocity resolution of $2$~km~s$^{-1}$ to $\approx100$~km~s$^{-1}$). Compared to normal star-forming galaxies, the velocity dispersions are elevated by $\approx0.15$~dex ($\approx40\%$) at a given mass surface density. Whilst there is a well-defined relationship between $\Sigma$ and $\sigma_\mathrm{EW}$ within each galaxy, the scatter of the zero-points leads to an overall weaker correlation when considering the full galaxy sample.
    
    \item The molecular gas of our sample ETGs is typically not in virial equilibrium, but rather is significantly super-virial (with a median virial parameter $\abeamvir=4.2^{+7.3}_{-3.2}$ at $60$~pc resolution, after weighting each galaxy evenly). There is a slight systematic trend with resolution, whereby $\abeamvir$ increases with worsening physical resolution, likely due to beam smearing of unresolved velocity gradients.
    
    \item The molecular gas of our sample ETGs spans two orders of magnitude in turbulent pressure $\pbeamturb$ (mainly reflecting the spread in $\Sigma$), and the median $\pbeamturb$ ($1.1^{+24.2}_{-1.0}\times10^7$ at $60$~pc resolution, again weighting each galaxy evenly) is significantly higher than that of the gas in star-forming discs.  $\pbeamturb$ in ETGs is thus more similar to that in the centres of spiral galaxies, or indeed that of the Milky Way Central Molecular Zone (CMZ) and the central kiloparsecs of simulated bulge-dominated galaxies. Again, there is some dependence of this parameter on the physical scale of the measurements.
    
    \item The molecular gas of our sample ETGs reveals possible anti-correlations of $\abeamvir$ and $\pbeamturb$ with the galaxy-integrated star-formation rate. Likely, the unusual conditions of the molecular gas are inhibiting star formation. The dynamical equilibrium pressures $\pde$ of our sample ETGs are generally significantly higher than $\pbeamturb$, and they become more dominant at larger galactocentric radii. This appears at odds with the high $\abeamvir$ measured, but likely indicates that large-scale dynamical forces, such as shear and tides, that are driven by the gravitational potentials and are ignored here, play a vital role in regulating the gas properties of ETGs.
\end{enumerate}

This work highlights that the molecular gas, the raw fuel for star formation, is strongly affected by its galactic environment. It is clear that the simple presence of molecular gas is not sufficient for star formation to switch on, and a careful balance of the internal gas properties and the external dynamics of the galaxy regulates the star-forming properties of clouds. As increasingly high resolution measurements of molecular gas are achieved with ALMA, extending observations beyond the galaxy main sequence will be vital to acquire a holistic overview of molecular gas properties and variations across the Universe. The molecular gas properties of both ultra-luminous infrared galaxies and ETGs are relatively poorly explored compared to those of main sequence galaxies, and a large sample of extragalactic CMZs will be vital to understand the most extreme bursts and dearths of star formation observed across galaxies.

\section*{Acknowledgements}

ALMA is a partnership of ESO (representing its member states), NSF (USA) and NINS (Japan), together with NRC (Canada), MOST and ASIAA (Taiwan), and KASI (Republic of Korea), in cooperation with the Republic of Chile. The Joint ALMA Observatory is operated by ESO, AUI/NRAO and NAOJ.

%%%%%%%%%%%%%%%%%%%%%%%%%%%%%%%%%%%%%%%%%%%%%%%%%%
\section*{Data Availability}

The raw data used in this study are all publicly available at \url{https://almascience.eso.org/asax/}. This paper makes use of the following ALMA data. For NGC~0383: ADS/JAO.ALMA\#2012.1.01092.S, ADS/JAO.ALMA\#2015.1.00419.S, ADS/JAO.ALMA\#2016.1.00437.S and ADS/JAO.ALMA\#2016.2.00053.S. For NGC~1574: ADS/JAO.ALMA\#2015.1.00419.S and ADS/JAO.ALMA\#2016.2.00053.S. For NGC~0524: ADS/JAO.ALMA\#2015.1.00466.S, ADS/JAO.ALMA\#2016.2.00053.S and ADS/JAO.ALMA\#2017.1.00391.S. For NGC~3607: ADS/JAO.ALMA\#2015.1.00598.S and ADS/JAO.ALMA\#2016.2.00053.S. For NGC~4429: ADS/JAO.ALMA\#2013.1.00493.S. For NGC~4435: ADS/JAO.ALMA\#2015.1.00598.S and ADS/JAO.ALMA\#2016.2.00053.S. For NGC~4697: ADS/JAO.ALMA\#2015.1.00598.S. Reduced cubes and maps are available at \url{https://www.canfar.net/storage/vault/list/AstroDataCitationDOI/CISTI.CANFAR/23.0016}. The scripts underlying this study are available at \url{https://github.com/thomaswilliamsastro/wisdom_px_gmc}, and the PHANGS-ALMA pipeline keys for reduction can be found at \url{https://github.com/thomaswilliamsastro/wisdom_alma_reduction}.

% The inclusion of a Data Availability Statement is a requirement for articles published in MNRAS. Data Availability Statements provide a standardised format for readers to understand the availability of data underlying the research results described in the article. The statement may refer to original data generated in the course of the study or to third-party data analysed in the article. The statement should describe and provide means of access, where possible, by linking to the data or providing the required accession numbers for the relevant databases or DOIs.


\bibliographystyle{mnras}
\bibliography{bibliography}

%%%%%%%%%%%%%%%%%%%%%%%%%%%%%%%%%%%%%%%%%%%%%%%%%%

%%%%%%%%%%%%%%%%% APPENDICES %%%%%%%%%%%%%%%%%%%%%

\appendix

\section{The $\sigma_\mathrm{EW}$ -- $\Sigma$ relations of individual galaxies}\label{app:sigma_sigma}

Here, we show a figure analogous to Figure~\ref{fig:sigma_sigma_ngc4429} for each sample galaxy.

\begin{figure*}
	\includegraphics[width=\textwidth]{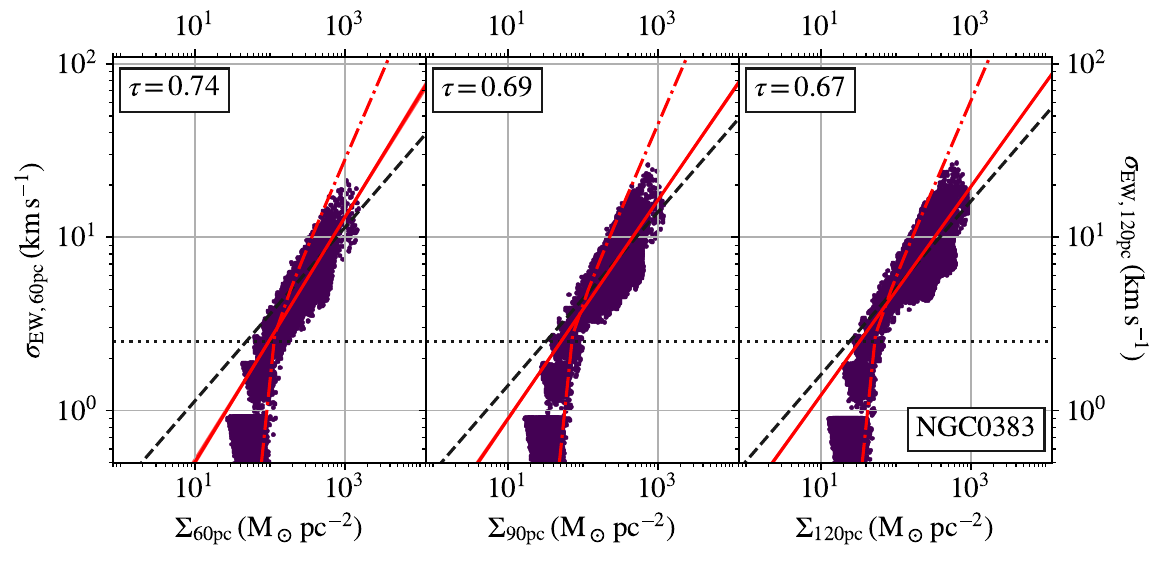}
    \caption{As Figure \ref{fig:sigma_sigma_ngc4429}, but for NGC~0383.}
    \label{fig:sigma_sigma_ngc0383}
\end{figure*}

\begin{figure*}
	\includegraphics[width=\textwidth]{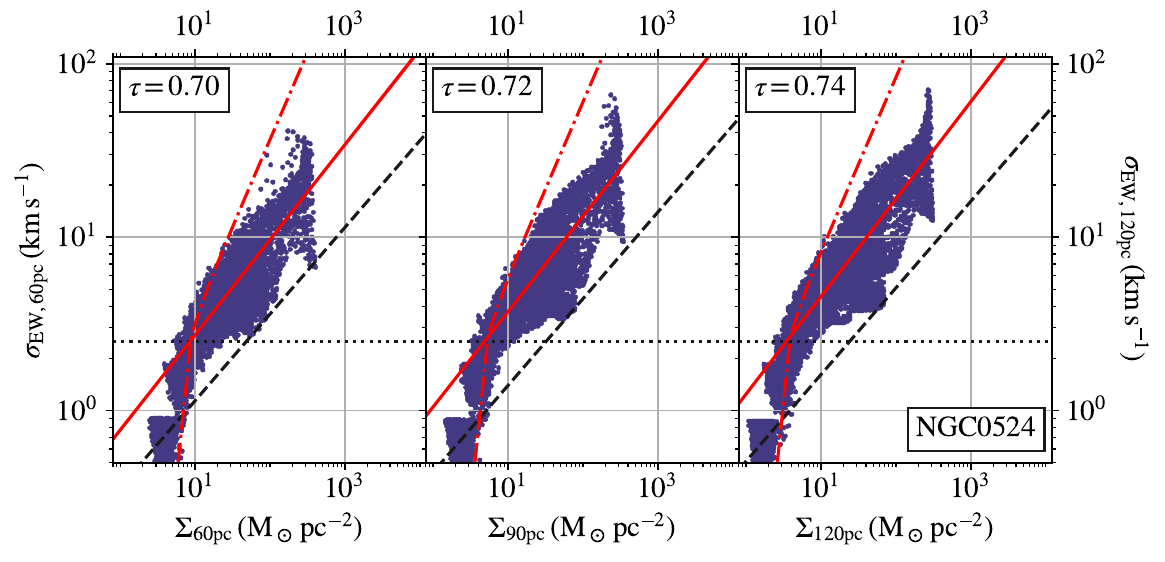}
    \caption{As Figure \ref{fig:sigma_sigma_ngc4429}, but for NGC~0524.}
    \label{fig:sigma_sigma_ngc0524}
\end{figure*}

\begin{figure*}
	\includegraphics[width=\textwidth]{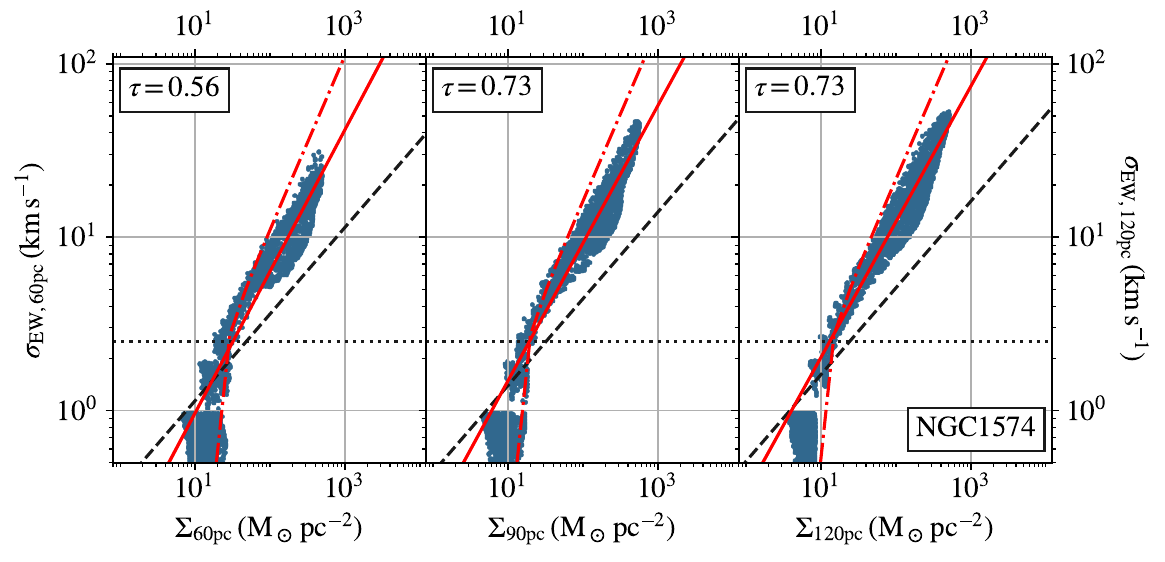}
    \caption{As Figure \ref{fig:sigma_sigma_ngc4429}, but for NGC~1574.}
    \label{fig:sigma_sigma_ngc1574}
\end{figure*}

\begin{figure*}
	\includegraphics[width=\textwidth]{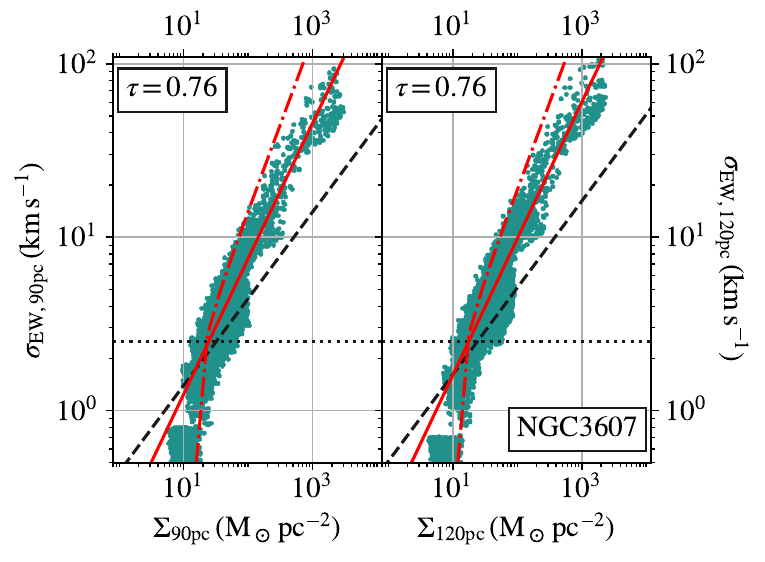}
    \caption{As Figure \ref{fig:sigma_sigma_ngc4429}, but for NGC~3607.}
    \label{fig:sigma_sigma_ngc3607}
\end{figure*}

\begin{figure*}
	\includegraphics[width=\textwidth]{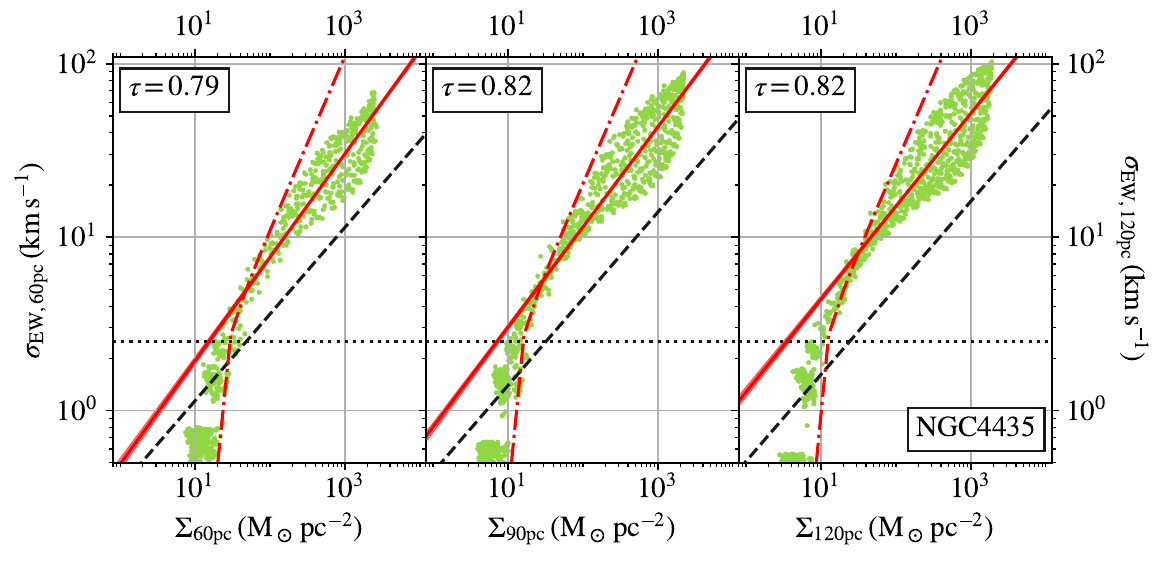}
    \caption{As Figure \ref{fig:sigma_sigma_ngc4429}, but for NGC~4435.}
    \label{fig:sigma_sigma_ngc4435}
\end{figure*}

\begin{figure*}
	\includegraphics[width=\textwidth]{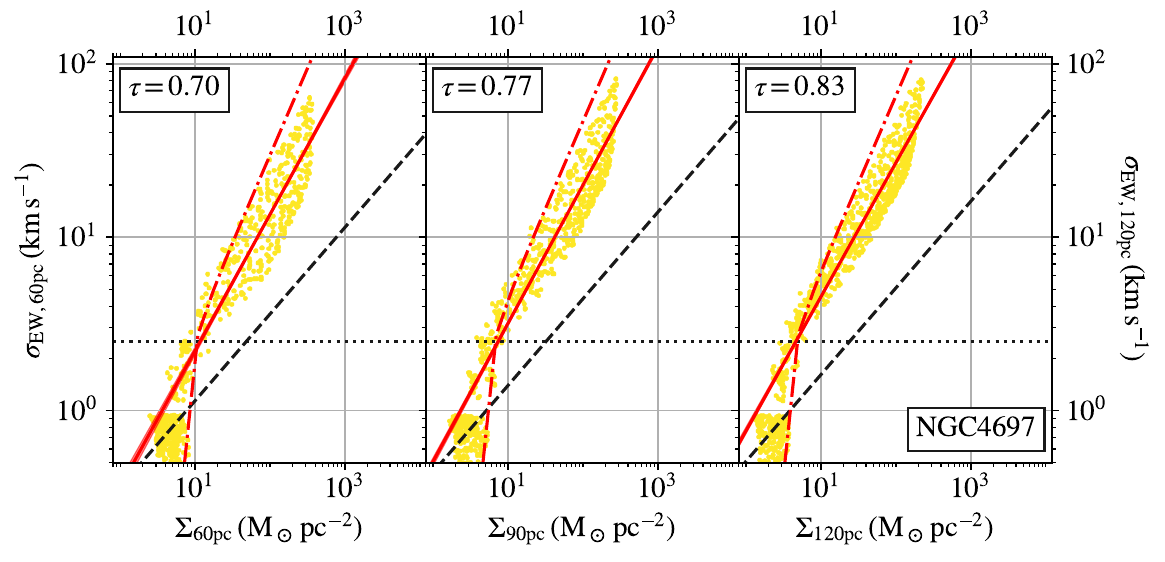}
    \caption{As Figure \ref{fig:sigma_sigma_ngc4429}, but for NGC~4697.}
    \label{fig:sigma_sigma_ngc4697}
\end{figure*}

\section{The $\abeamvir$ distributions of individual galaxies}\label{app:avir}

Here, we show a figure analogous to Figure~\ref{fig:virial_parameter_ngc4429} for each sample galaxy.

\begin{figure*}
	\includegraphics[width=\textwidth]{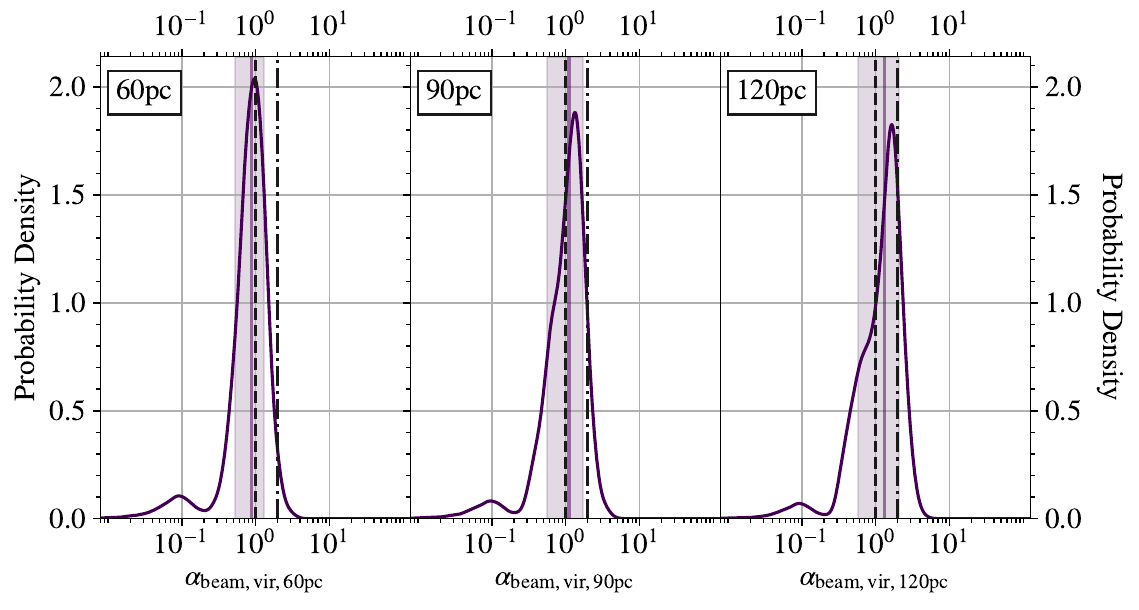}
    \caption{As Figure \ref{fig:virial_parameter_ngc4429}, but for NGC~0383.}
    \label{fig:virial_parameter_ngc0383}
\end{figure*}

\begin{figure*}
	\includegraphics[width=\textwidth]{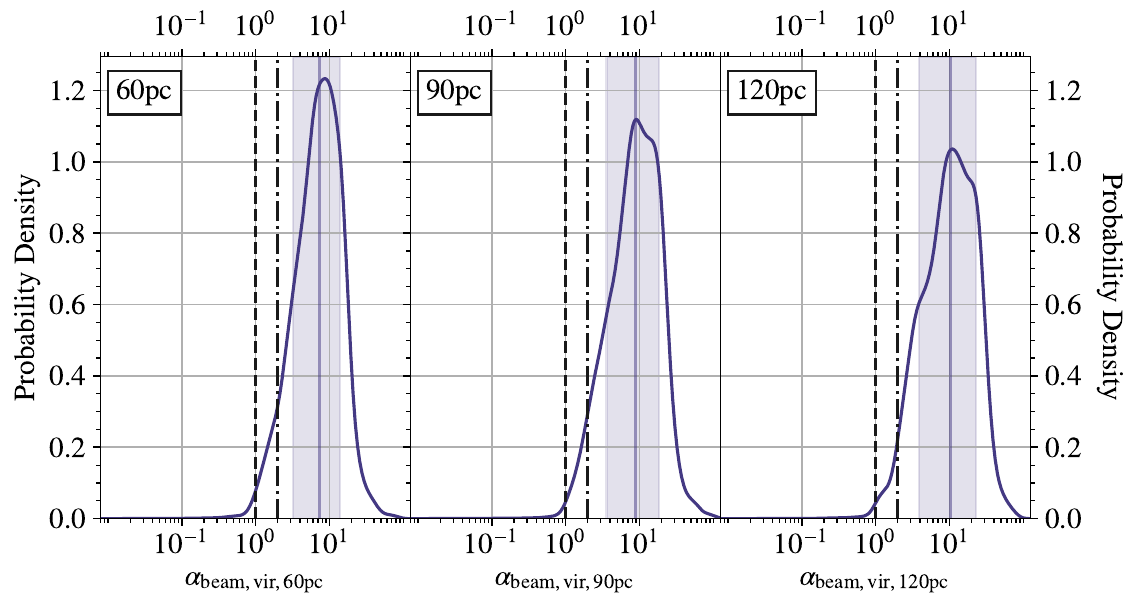}
    \caption{As Figure \ref{fig:virial_parameter_ngc4429}, but for NGC~0524.}
    \label{fig:virial_parameter_ngc0524}
\end{figure*}

\begin{figure*}
	\includegraphics[width=\textwidth]{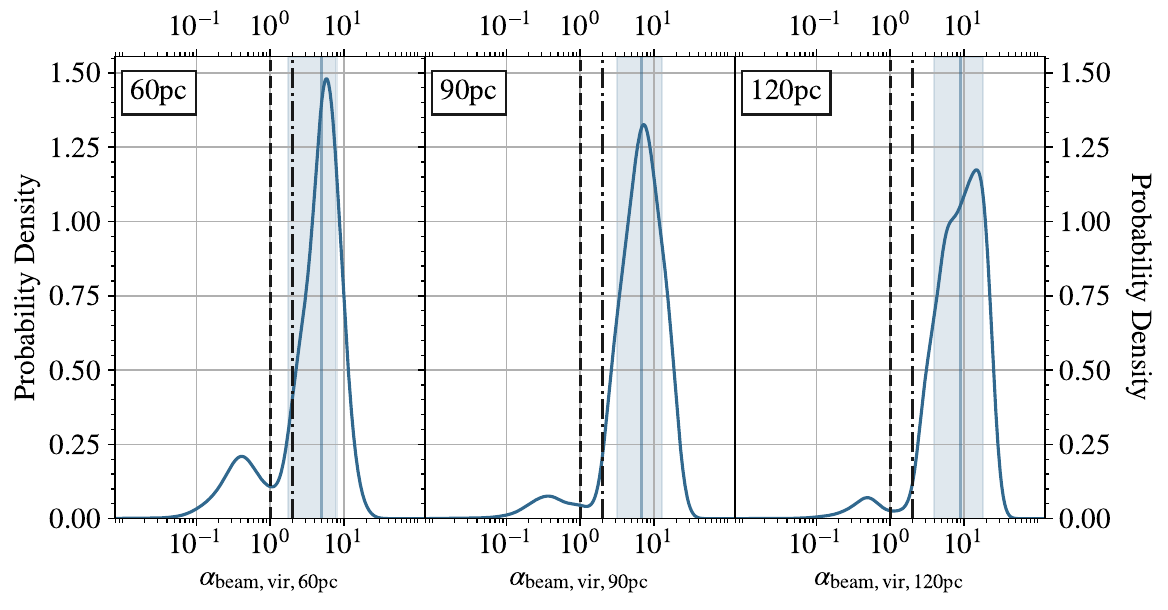}
    \caption{As Figure \ref{fig:virial_parameter_ngc4429}, but for NGC~1574.}
    \label{fig:virial_parameter_ngc1574}
\end{figure*}

\begin{figure*}
	\includegraphics[width=\textwidth]{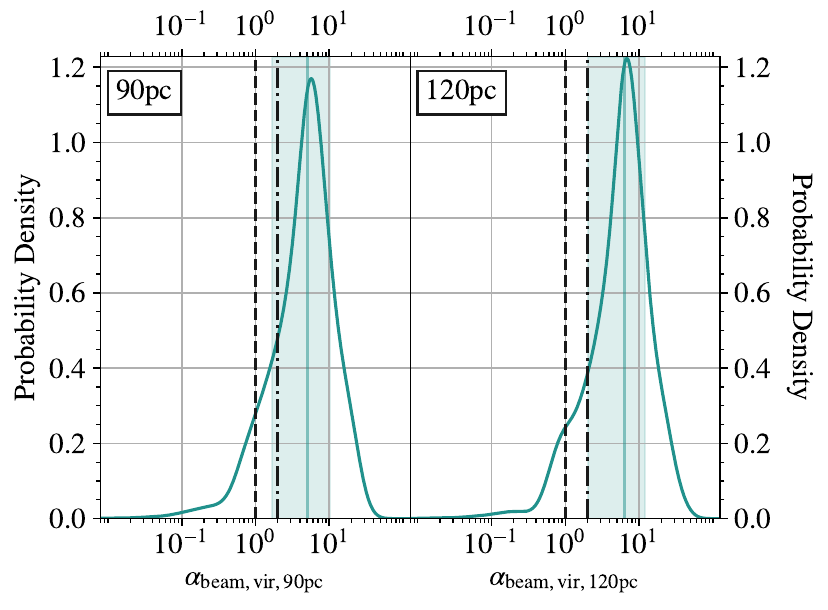}
    \caption{As Figure \ref{fig:virial_parameter_ngc4429}, but for NGC~3607.}
    \label{fig:virial_parameter_ngc3607}
\end{figure*}

\begin{figure*}
	\includegraphics[width=\textwidth]{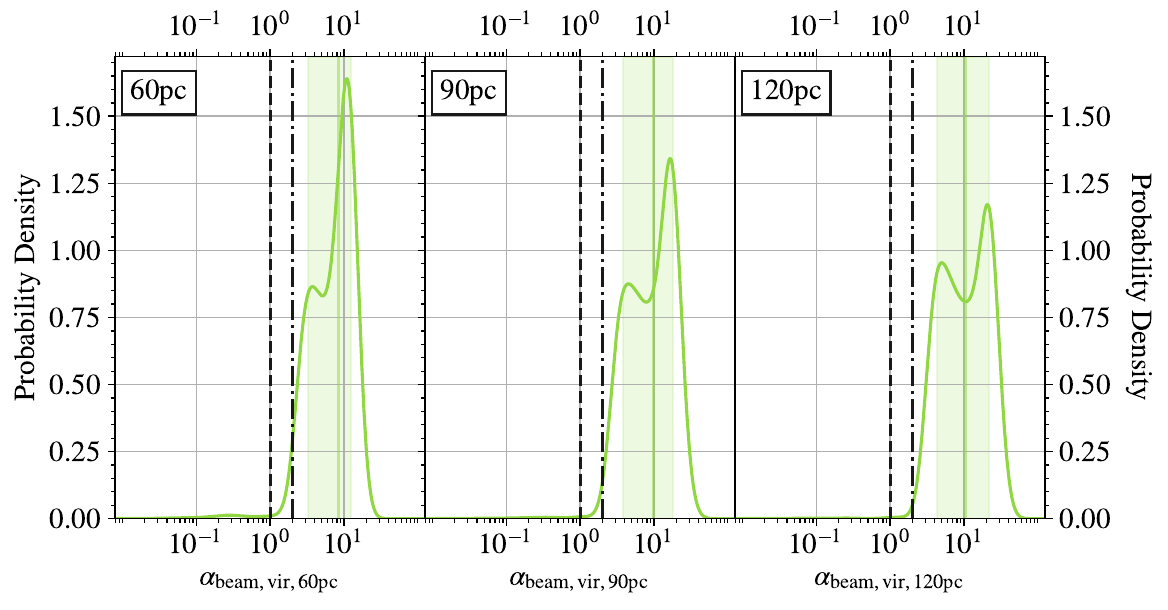}
    \caption{As Figure \ref{fig:virial_parameter_ngc4429}, but for NGC~4435.}
    \label{fig:virial_parameter_ngc4435}
\end{figure*}

\begin{figure*}
	\includegraphics[width=\textwidth]{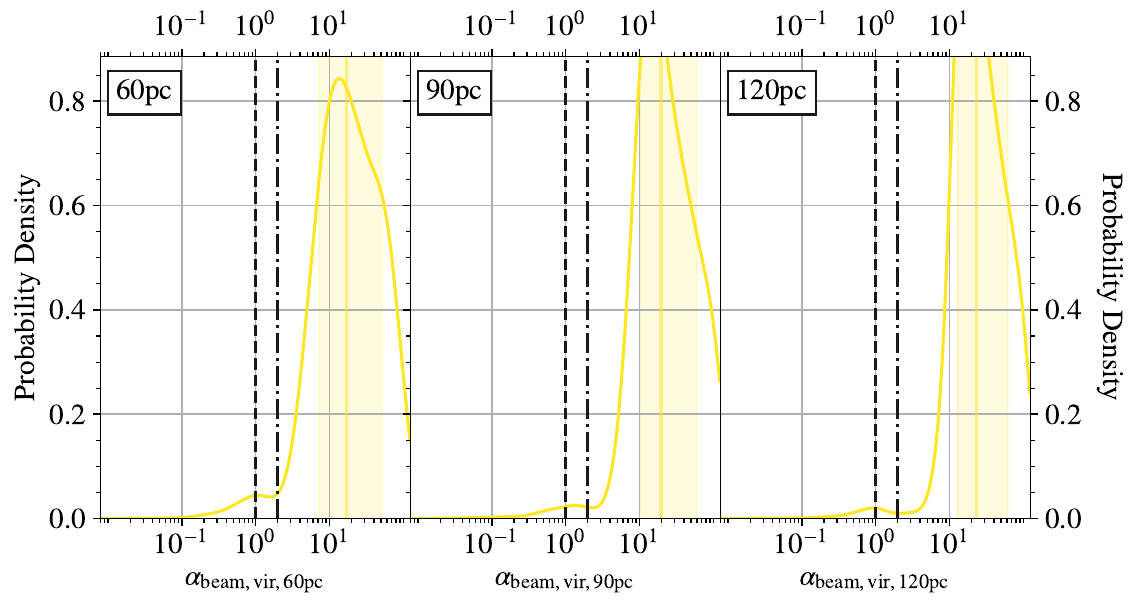}
    \caption{As Figure \ref{fig:virial_parameter_ngc4429}, but for NGC~4697.}
    \label{fig:virial_parameter_ngc4697}
\end{figure*}

\section{The $\pbeamturb$ distributions of individual galaxies}\label{app:pturb}

Here, we show a figure analogous to Figure~\ref{fig:pturb_ngc4429} for each sample galaxy.

\begin{figure*}
	\includegraphics[width=\textwidth]{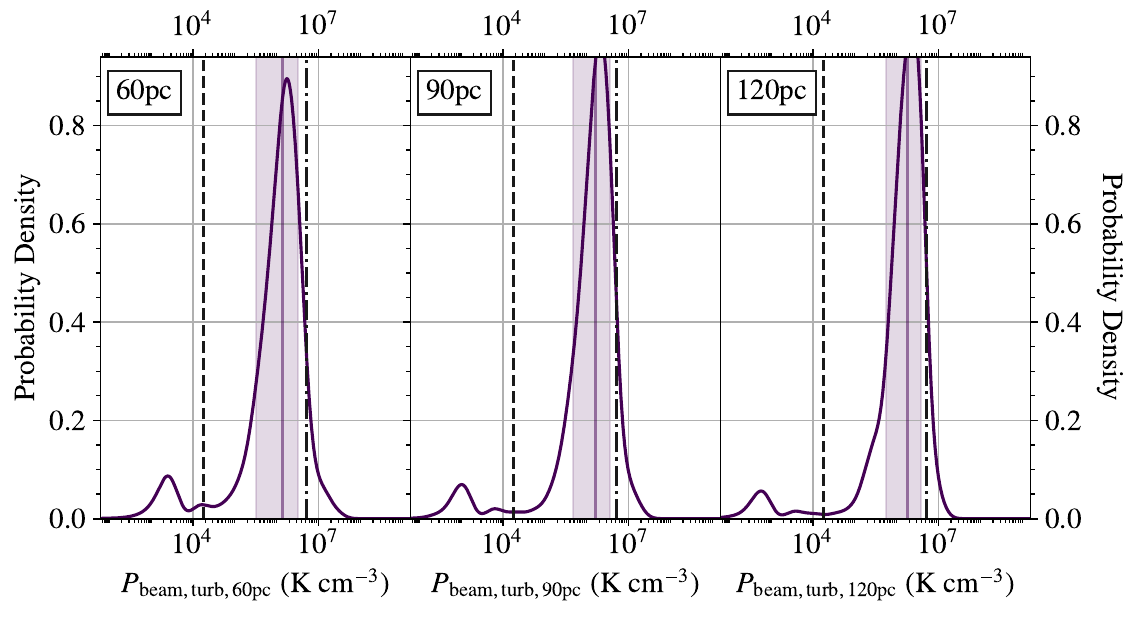}
    \caption{As Figure \ref{fig:pturb_ngc4429}, but for NGC~0383.}
    \label{fig:pturb_ngc0383}
\end{figure*}

\begin{figure*}
	\includegraphics[width=\textwidth]{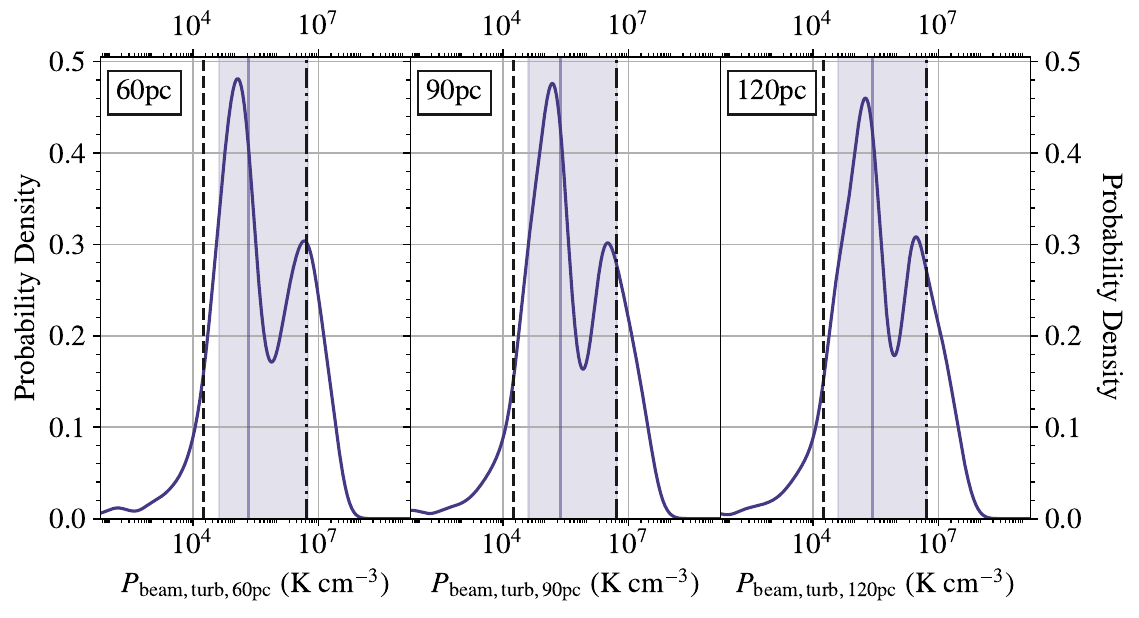}
    \caption{As Figure \ref{fig:pturb_ngc4429}, but for NGC~0524.}
    \label{fig:pturb_ngc0524}
\end{figure*}

\begin{figure*}
	\includegraphics[width=\textwidth]{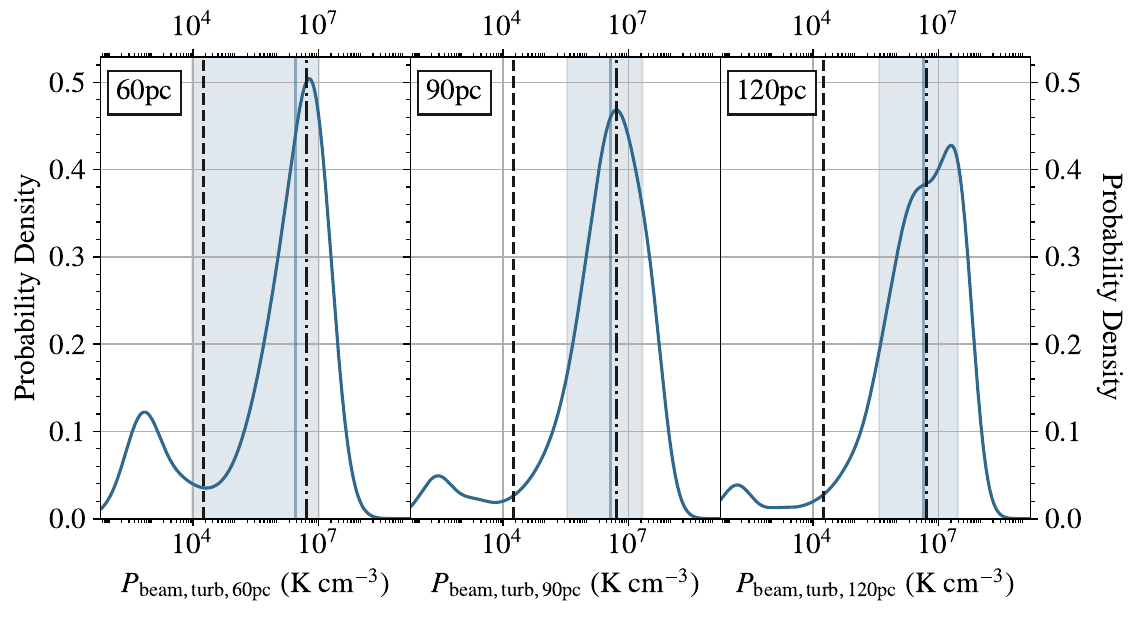}
    \caption{As Figure \ref{fig:pturb_ngc4429}, but for NGC~1574.}
    \label{fig:pturb_ngc1574}
\end{figure*}

\begin{figure*}
	\includegraphics[width=\textwidth]{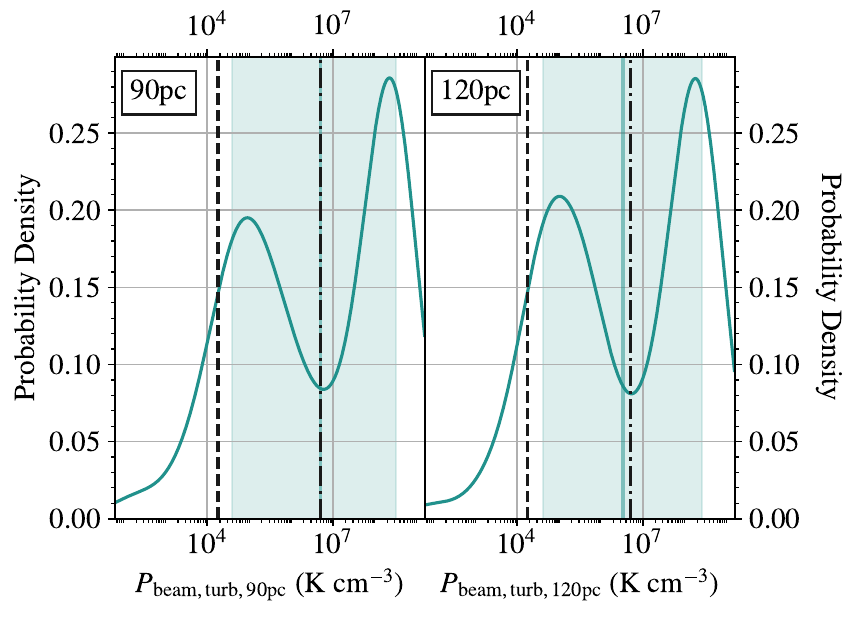}
    \caption{As Figure \ref{fig:pturb_ngc4429}, but for NGC~3607.}
    \label{fig:pturb_ngc3607}
\end{figure*}

\begin{figure*}
	\includegraphics[width=\textwidth]{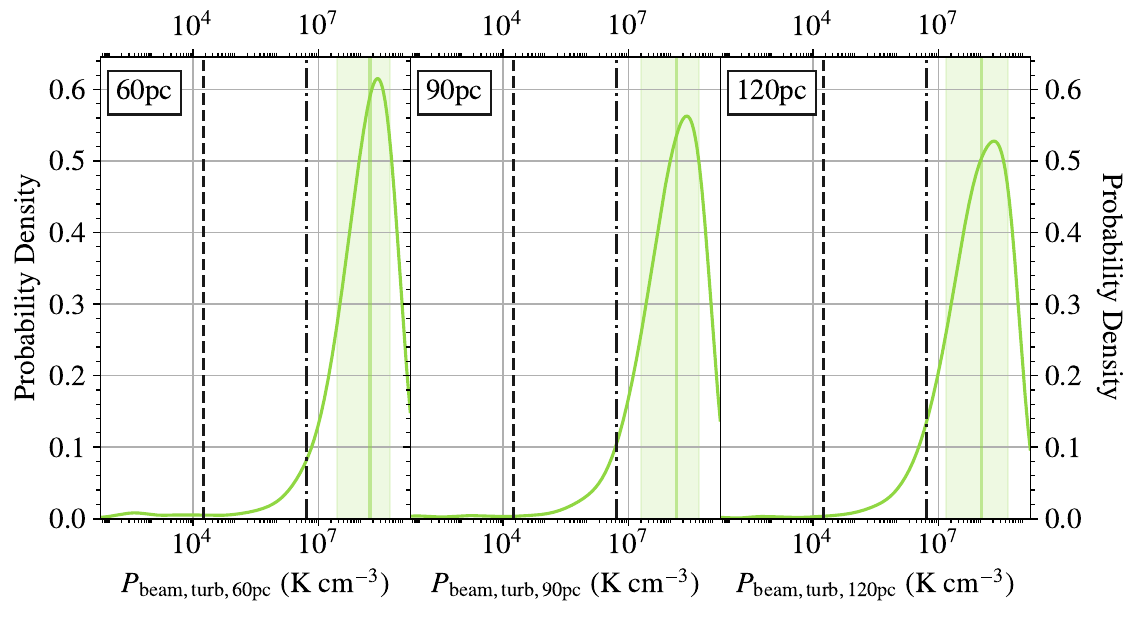}
    \caption{As Figure \ref{fig:pturb_ngc4429}, but for NGC~4435.}
    \label{fig:pturb_ngc4435}
\end{figure*}

\begin{figure*}
	\includegraphics[width=\textwidth]{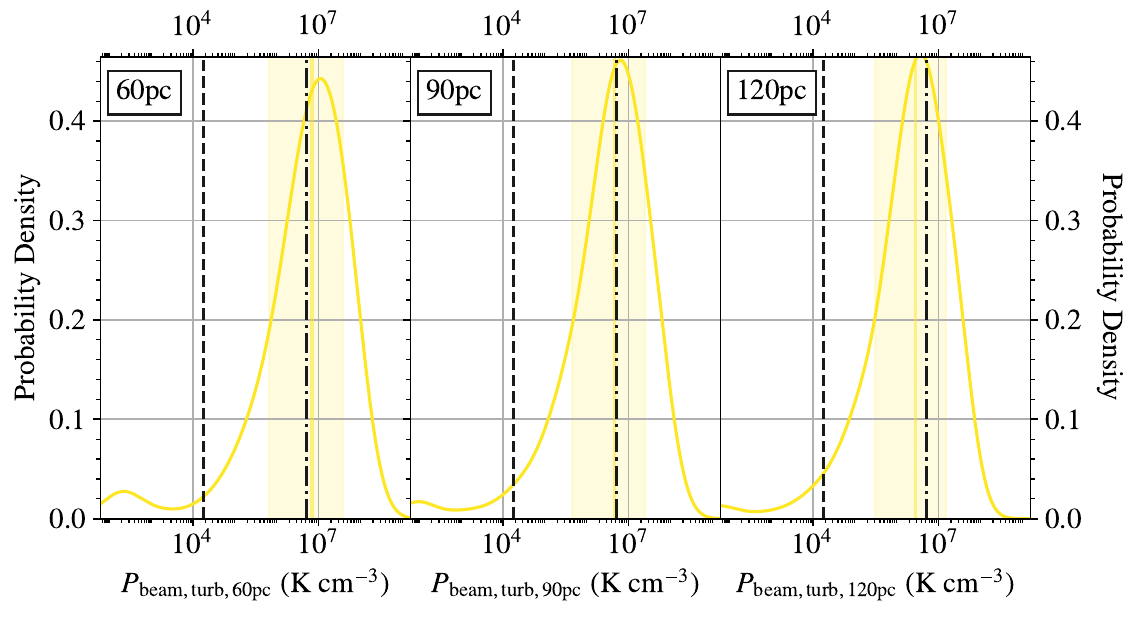}
    \caption{As Figure \ref{fig:pturb_ngc4429}, but for NGC~4697.}
    \label{fig:pturb_ngc4697}
\end{figure*}

\section{The $\pbeamturb/\pde$ maps of individual galaxies}\label{app:pressure_ratio}

Here, we show a figure analogous to Figure~\ref{fig:pressure_ratio_ngc4429} for each sample galaxy.

\begin{figure*}
	\includegraphics[width=\textwidth]{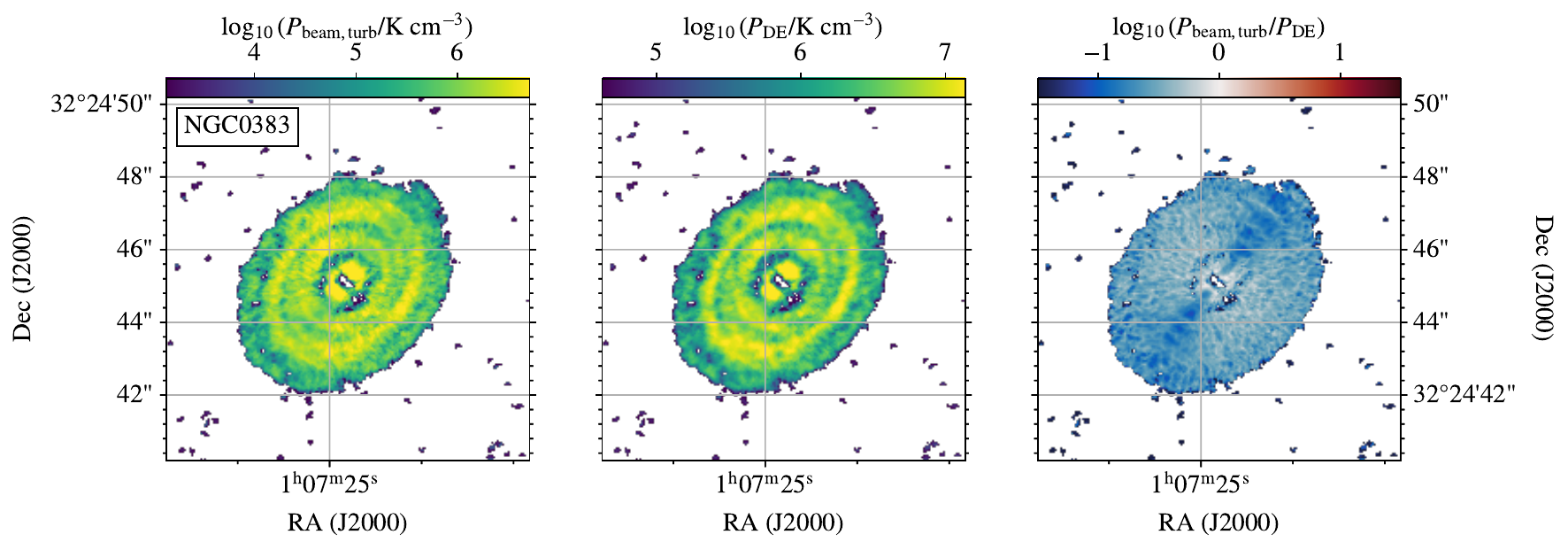}
    \caption{As Figure \ref{fig:pressure_ratio_ngc4429}, but for NGC~0383.}
    \label{fig:pressure_ratio_ngc0383}
\end{figure*}

\begin{figure*}
	\includegraphics[width=\textwidth]{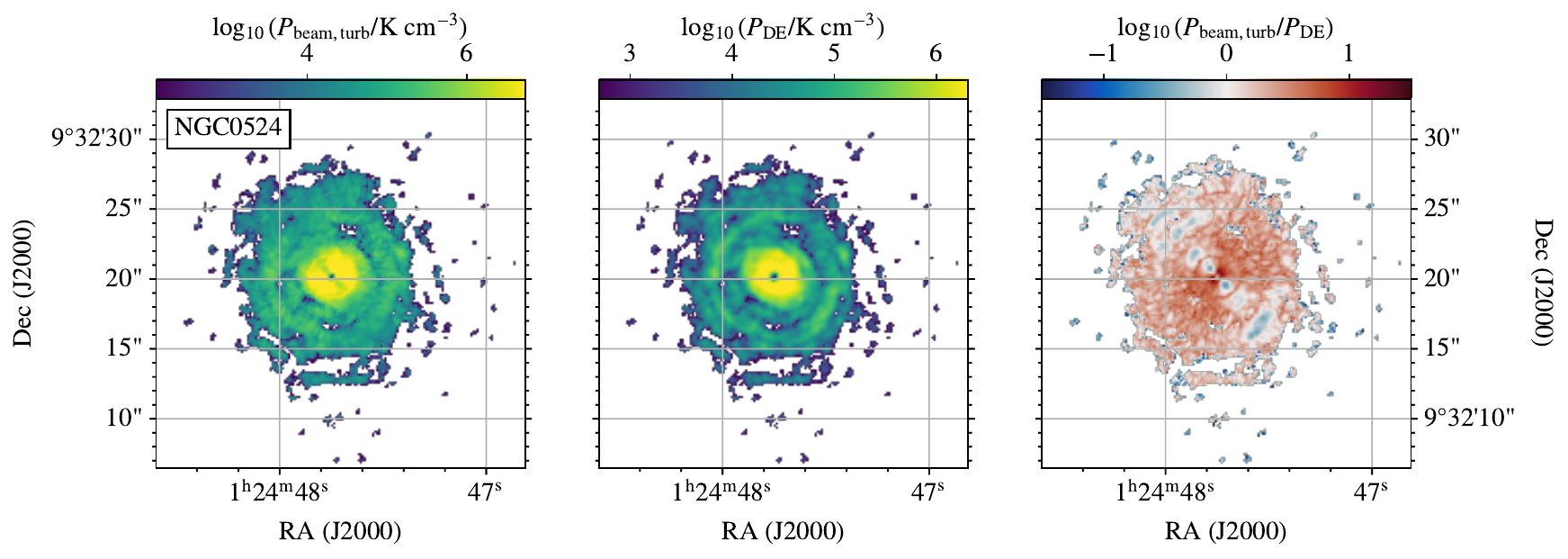}
    \caption{As Figure \ref{fig:pressure_ratio_ngc4429}, but for NGC~0524.}
    \label{fig:pressure_ratio_ngc0524}
\end{figure*}

\begin{figure*}
	\includegraphics[width=\textwidth]{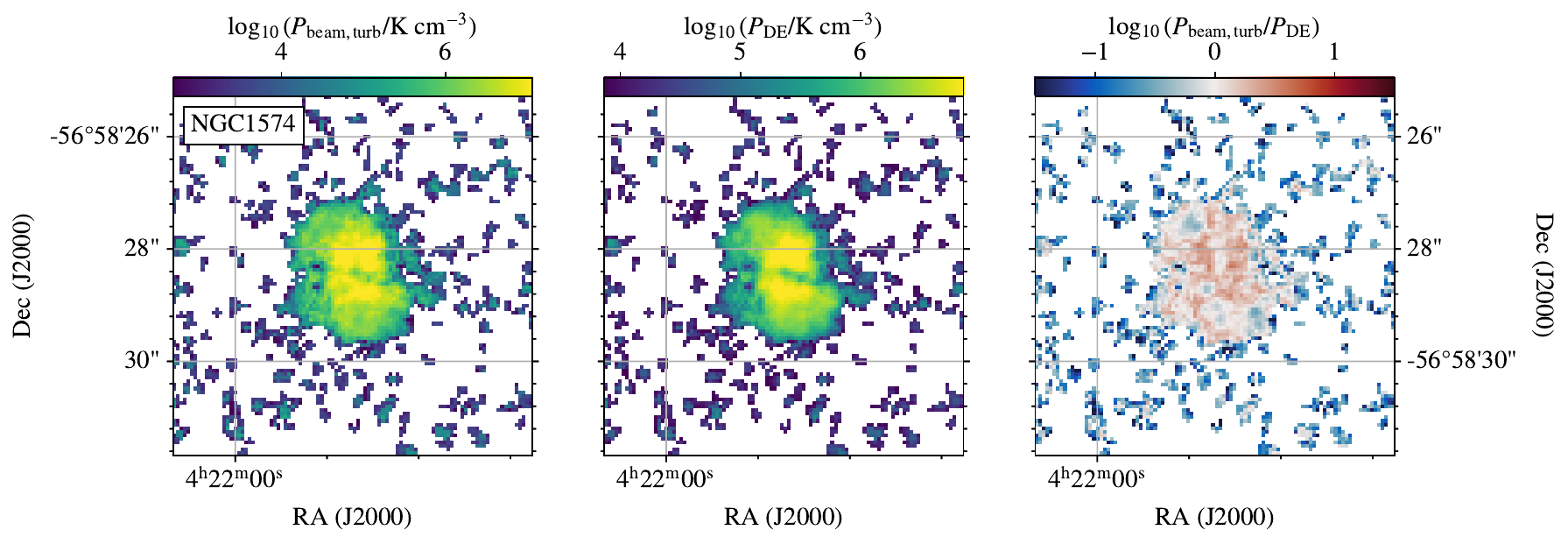}
    \caption{As Figure \ref{fig:pressure_ratio_ngc4429}, but for NGC~1574.}
    \label{fig:pressure_ratio_ngc1574}
\end{figure*}

\begin{figure*}
	\includegraphics[width=\textwidth]{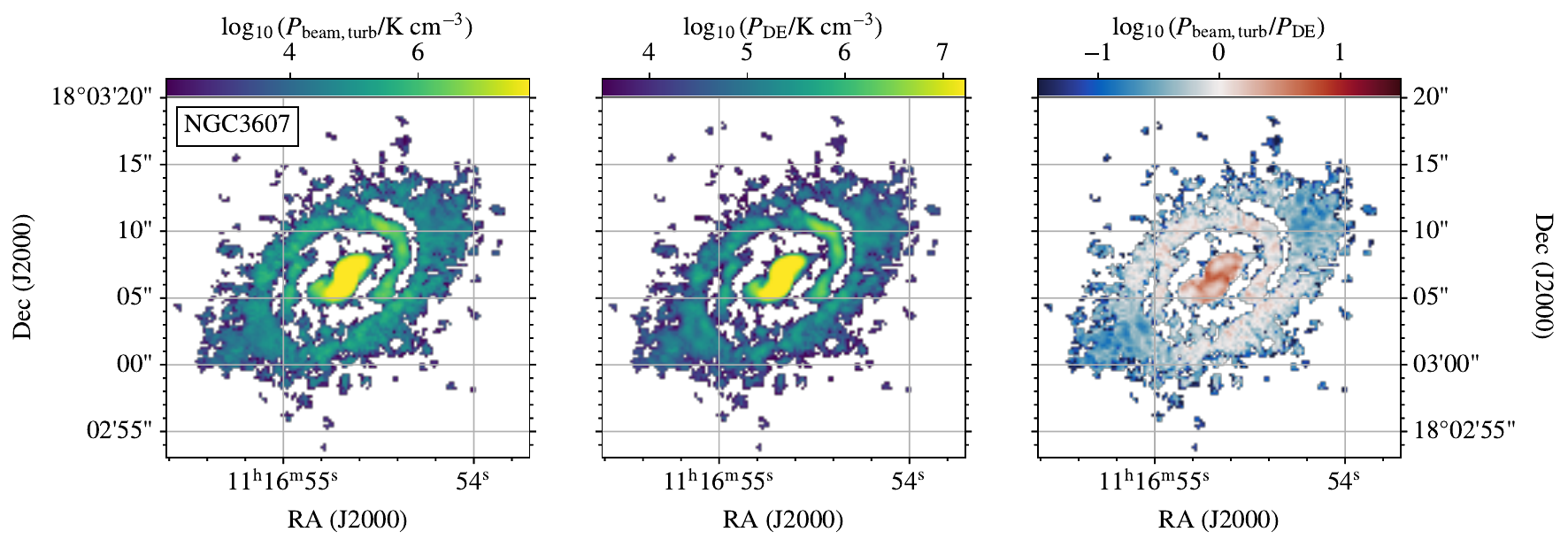}
    \caption{As Figure \ref{fig:pressure_ratio_ngc4429}, but for NGC~3607.}
    \label{fig:pressure_ratio_ngc3607}
\end{figure*}

\begin{figure*}
	\includegraphics[width=\textwidth]{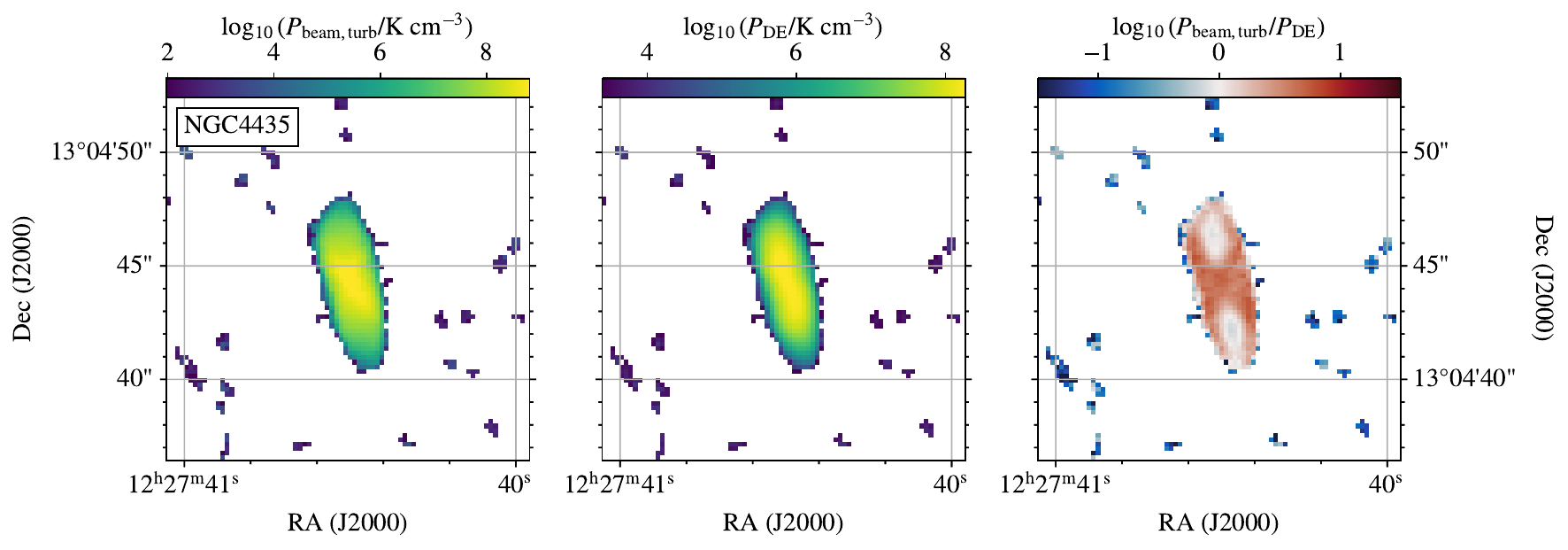}
    \caption{As Figure \ref{fig:pressure_ratio_ngc4429}, but for NGC~4435.}
    \label{fig:pressure_ratio_ngc4435}
\end{figure*}

\begin{figure*}
	\includegraphics[width=\textwidth]{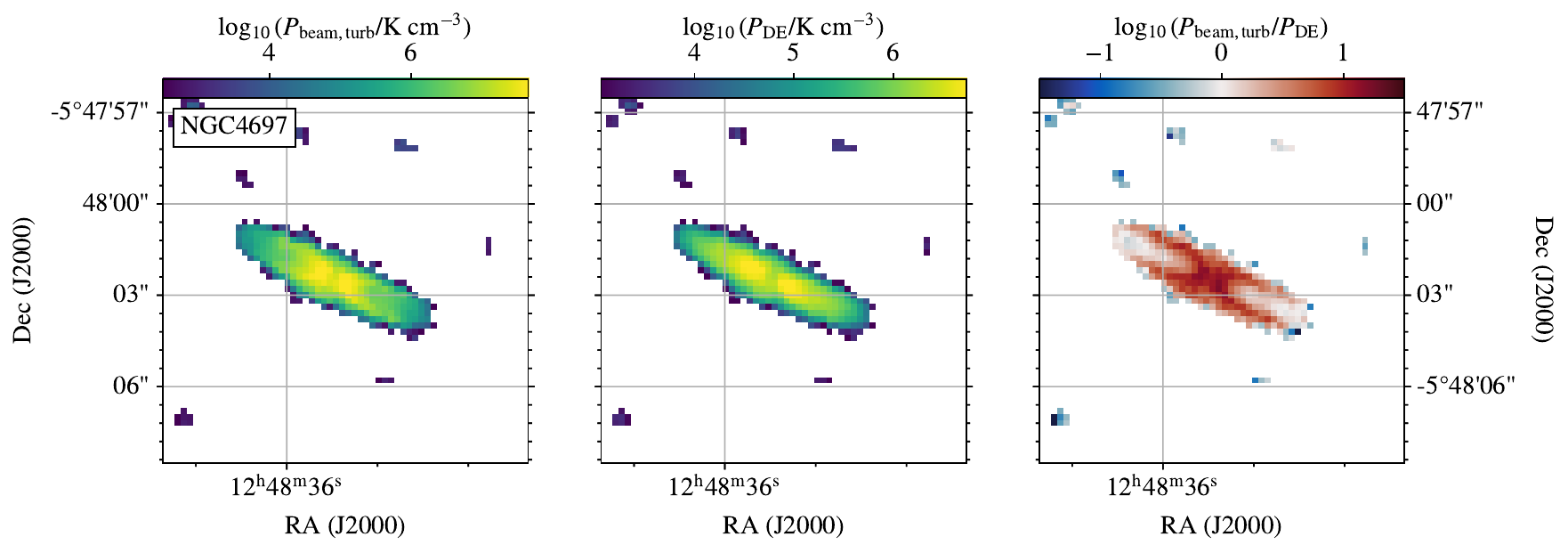}
    \caption{As Figure \ref{fig:pressure_ratio_ngc4429}, but for NGC~4697.}
    \label{fig:pressure_ratio_ngc4697}
\end{figure*}

\section{$\abeamvir$ versus SFR/SFE at different resolutions}\label{app:avir_sfr}

Here, we show a figure analogous to Figure~\ref{fig:alpha_vir_sfr} for each sample galaxy.

\begin{figure*}
	\includegraphics[width=\textwidth]{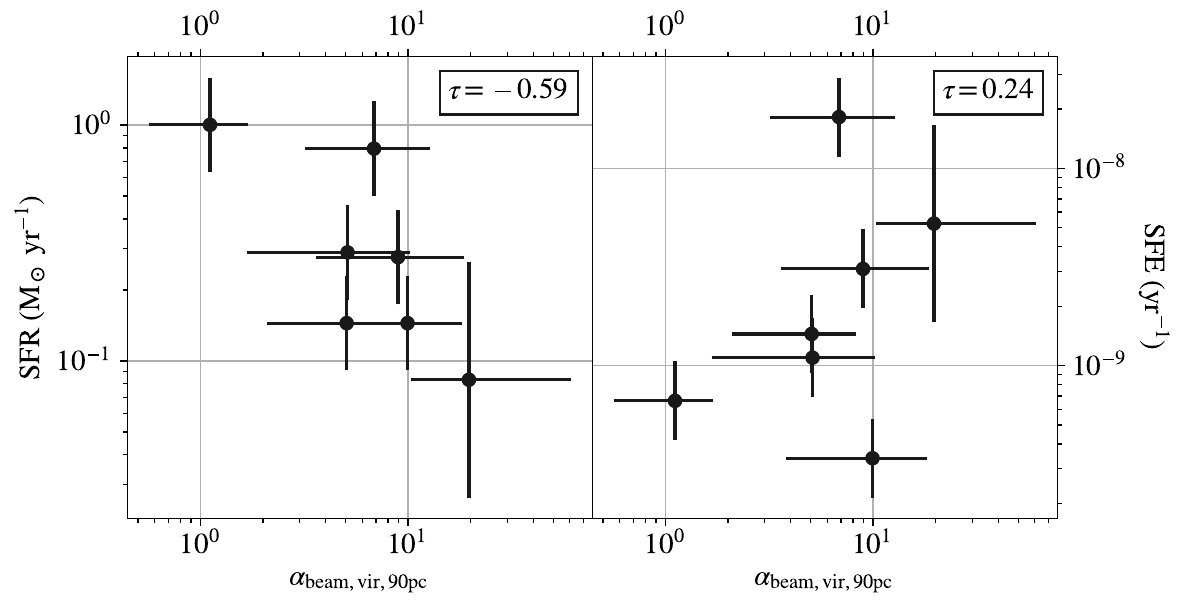}
    \caption{As Figure \ref{fig:alpha_vir_sfr}, but at 90~pc resolution.}
    \label{fig:alpha_vir_sfr_90}
\end{figure*}

\begin{figure*}
	\includegraphics[width=\textwidth]{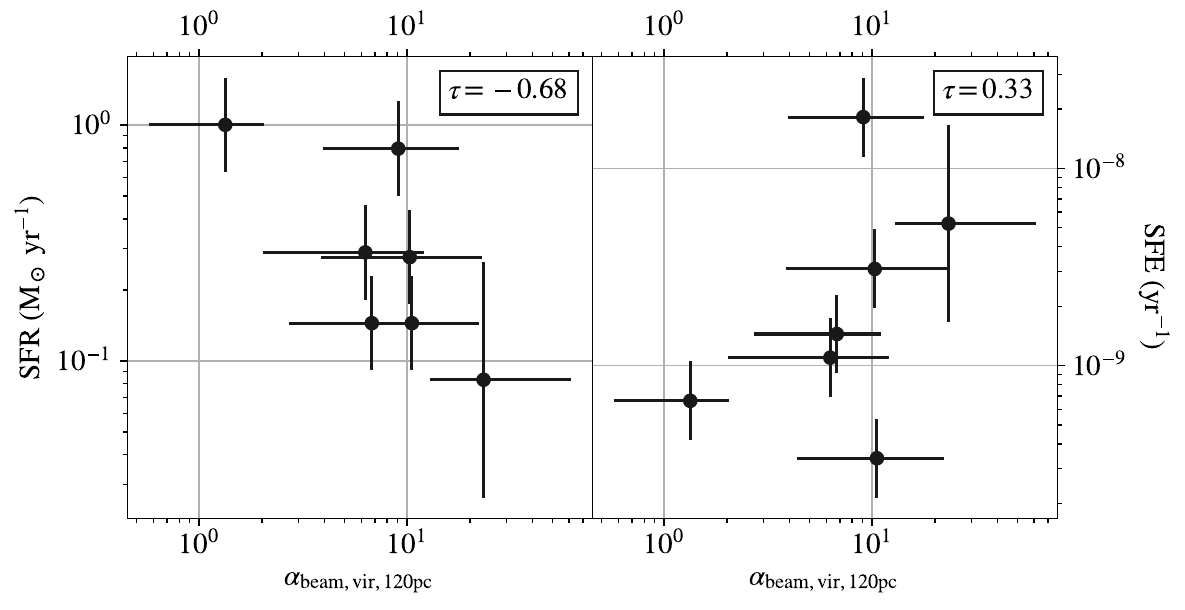}
    \caption{As Figure \ref{fig:alpha_vir_sfr}, but at 120~pc resolution.}
    \label{fig:alpha_vir_sfr_120}
\end{figure*}

\section{$\pbeamturb$ versus SFR/SFE at different resolutions}\label{app:pturb_sfr}

Here, we show a figure analogous to Figure~\ref{fig:pressure_sfr} for each sample galaxy.

\begin{figure*}
	\includegraphics[width=\textwidth]{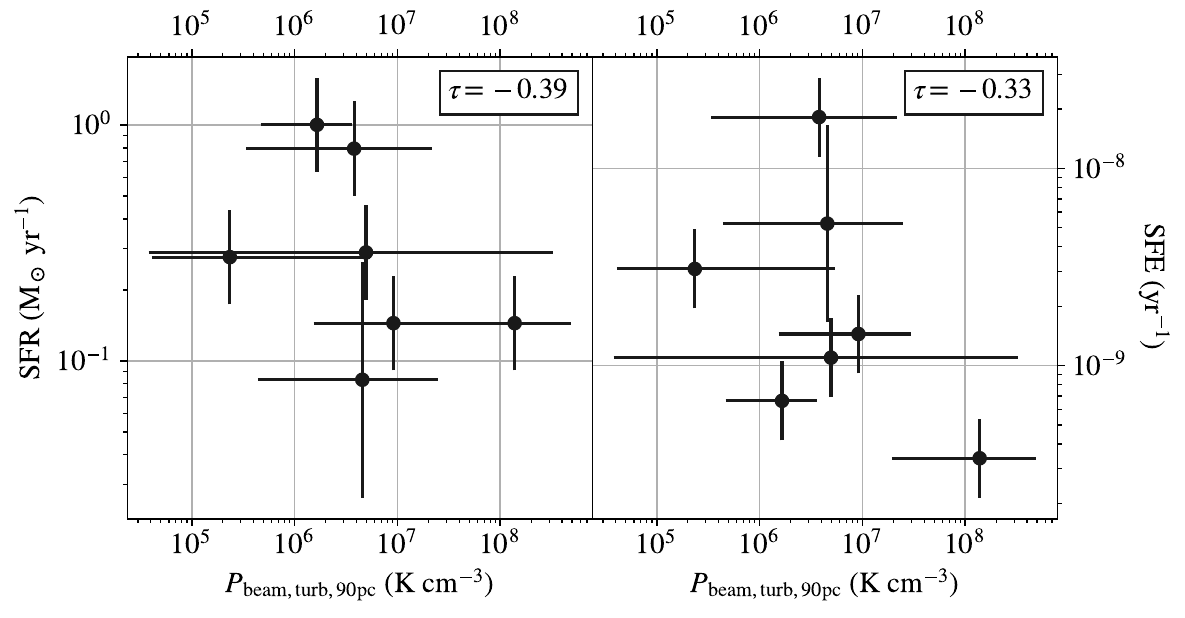}
    \caption{As Figure \ref{fig:pressure_sfr}, but at 90~pc resolution.}
    \label{fig:pressure_sfr_90}
\end{figure*}

\begin{figure*}
	\includegraphics[width=\textwidth]{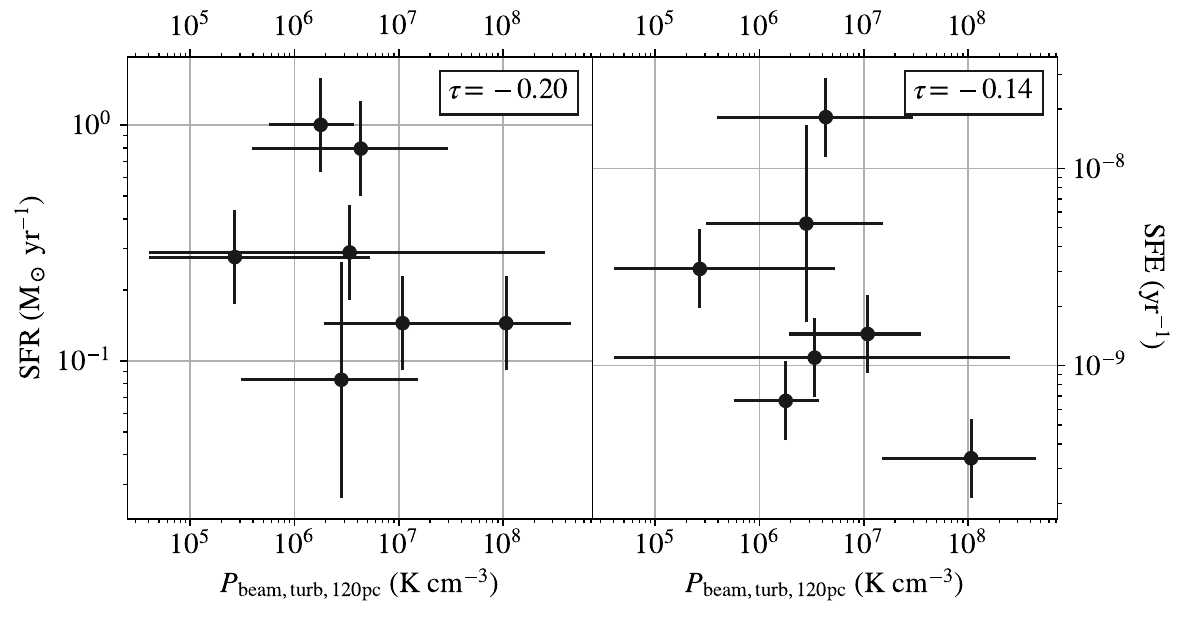}
    \caption{As Figure \ref{fig:pressure_sfr}, but at 120~pc resolution.}
    \label{fig:pressure_sfr_120}
\end{figure*}

% If you want to present additional material which would interrupt the flow of the main paper,
% it can be placed in an Appendix which appears after the list of references.

%%%%%%%%%%%%%%%%%%%%%%%%%%%%%%%%%%%%%%%%%%%%%%%%%%

% Don't change these lines
\bsp	% typesetting comment
\label{lastpage}
\end{document}